\documentclass[reprint,amsmath,amssymb,pra,aps,nofootinbib,superscriptaddress,floatfix]{revtex4-1}

\usepackage[T1]{fontenc}
\usepackage[utf8]{inputenc}
\usepackage{amsmath, amssymb}
\usepackage{bbold}
\usepackage{bm}
\usepackage{tipa}
\usepackage{subfigure}
\usepackage{blindtext}
\usepackage[dvipsnames]{xcolor}
\definecolor{dcyan}{RGB}{0,100,100}
\usepackage{graphicx}
\usepackage{tabularx}
\usepackage{textgreek}
\usepackage{xfrac}
\usepackage{units}
\usepackage{comment}
\usepackage{float}
\usepackage{placeins}
\usepackage{upgreek}
\usepackage{booktabs}  
\usepackage{threeparttable}
\usepackage{multirow}

\usepackage[hypertexnames=false]{hyperref}
\usepackage{xcolor}
\definecolor{green_cust}{RGB}{0,154,85}
\definecolor{red_cust}{RGB}{173,49,54}
\definecolor{blue_cust}{RGB}{0,103,148}
\hypersetup{colorlinks=true, linkcolor=red_cust, citecolor=green_cust, filecolor=blue_cust, urlcolor=blue_cust}

\makeatletter 
    
\renewcommand\onecolumngrid{
\do@columngrid{one}{\@ne}%
\def\set@footnotewidth{\onecolumngrid}
\def\footnoterule{\kern-6pt\hrule width 1.5in\kern6pt}%
}

\renewcommand\twocolumngrid{
        \def\footnoterule{
        \dimen@\skip\footins\divide\dimen@\thr@@
        \kern-\dimen@\hrule width.5in\kern\dimen@}
        \do@columngrid{mlt}{\tw@}
}%

\makeatother    

\usepackage{soul}

\newcommand{\Applied}{Department of Applied Physics, Stanford University, Stanford, CA}
\newcommand{\Physics}{Department of Physics, Stanford University, Stanford, CA}
\newcommand{\Montana}{Department of Physics, Montana State University, Bozeman, MT}
\newcommand{\StonyBrook}{Department of Physics and Astronomy, Stony Brook University, Stony Brook, NY}
\newcommand{\Harvard}{Department of Physics, Harvard University, Boston, MA}
\newcommand{\Chicago}{Department of Physics, The University of Chicago, Chicago, IL}

\newcommand{\Figref}[1]{Fig.~\hyperref[#1]{\ref{#1}}}

\begin{document}
\title{A cavity array microscope for parallel single-atom interfacing}



\author{Adam L. Shaw}\thanks{These authors contributed equally.}
\affiliation{\Physics}
\affiliation{\Applied}
\author{Anna Soper}\thanks{These authors contributed equally.}
\affiliation{\Applied}
\author{Danial Shadmany}\thanks{These authors contributed equally.}
\affiliation{\Physics}
\author{Aishwarya Kumar}
\affiliation{\StonyBrook}
\author{Lukas Palm}
\affiliation{\Chicago}
\author{Da-Yeon Koh}
\affiliation{\Physics}
\author{Vassilios Kaxiras}
\affiliation{\Harvard}
\author{Lavanya Taneja}
\affiliation{\Chicago}
\author{Matt Jaffe}
\affiliation{\Montana}
\author{David I. Schuster}
\affiliation{\Applied}
\author{Jonathan Simon}\email{jonsimon@stanford.edu}
\affiliation{\Physics}
\affiliation{\Applied}


\begin{abstract}
Neutral atom arrays and optical cavity QED systems have developed in parallel as central pillars of modern experimental quantum science~\cite{kaufman2021quantum,walther2006cavity,reiserer2015cavity}. While each platform has demonstrated exceptional capabilities—such as high-fidelity quantum logic~\cite{tsai2025benchmarking,evered2024highfidelity,peper2025spectroscopy,radnaev2025universal} in atom arrays, and strong light-matter coupling in cavities~\cite{birnbaum2005photon,haas2014entangled,barontini2015deterministic}—their combination holds promise for realizing fast and non-destructive atom measurement~\cite{wang2025ultrafast}, building large-scale quantum networks~\cite{huie2021multiplexed,li2024highrate,covey2023quantum,zhang2025proposal,sinclair2025faulttolerant,hahn2025deterministic}, and engineering hybrid atom-photon Hamiltonians~\cite{makin2008quantum,hartmann2008quantum,greentree2006quantum}. However, to date, experiments integrating the two platforms have been limited to spatially interfacing the entire atom array with one global cavity mode~\cite{deist2022mid,grinkemeyer2025errordetected,hu2025siteselective,hartung2024quantumnetwork,liu2023realization,dhordjevic2021entanglement}, a configuration that constrains addressability, parallelism, and scalability. Here we introduce the cavity array microscope, an experimental platform where each individual atom is strongly coupled to its own individual cavity across a two-dimensional array of over 40 modes. Our approach requires no nanophotonic elements~\cite{menon2024integrated,dhordjevic2021entanglement}, and instead uses a new free-space cavity geometry with intra-cavity lenses~\cite{shadmany2025cavity,mahler2019coupling} to realize above-unity peak cooperativity with micron-scale mode waists and spacings, compatible with typical atom array length scales while keeping atoms far from dielectric surfaces. We achieve homogeneous atom-cavity coupling, and show fast, non-destructive, parallel readout on millisecond timescales, including through a fiber array as a proof-of-principle for networking applications~\cite{li2025parallelized}. As an outlook, we realize a next-generation iteration of the platform with over 500 cavities and a nearly $10\times$ improvement in finesse. Our work unlocks the regime of many-cavity QED, and opens an unexplored frontier of large-scale quantum networking with atom arrays.

\end{abstract}
\maketitle

\subsection*{Introduction}
\label{sec:intro}

Neutral atom arrays have emerged as a promising platform for quantum information processing, having demonstrated two-qubit gate fidelities approaching 99.9\%~\cite{tsai2025benchmarking,evered2024highfidelity,peper2025spectroscopy,radnaev2025universal} and system sizes nearing 10,000 atoms~\cite{manetsch2025tweezer}. Scaling to the yet-larger systems believed to be necessary for useful, fault-tolerant applications~\cite{beverland2022assessing}, however, appears challenging in a single apparatus. This has spurred interest in the development of a modular quantum processing network~\cite{kimble2008quantum,rached2024benchmarking} composed of local atom array nodes interconnected by optical fibers~\cite{huie2021multiplexed,li2024highrate,covey2023quantum,sinclair2025faulttolerant,hahn2025deterministic,zhang2025proposal}. Central to this vision is the optical cavity, a platform in which multiple passes of confined light constructively interfere to form resonant modes that strongly enhance light-matter coupling~\cite{reiserer2015cavity,birnbaum2005photon,haas2014entangled,barontini2015deterministic}. Such strong coupling improves atom readout rates for detection and networking, and enables engineering of Hamiltonians for both quantum simulation~\cite{sauerwein2023engineering,hartmann2008quantum,young2024observing} and sensing~\cite{leroux2011implementation,hosten2016measurement}. 

\begin{figure}[t!]
	\centering
 	\includegraphics[width=82mm]{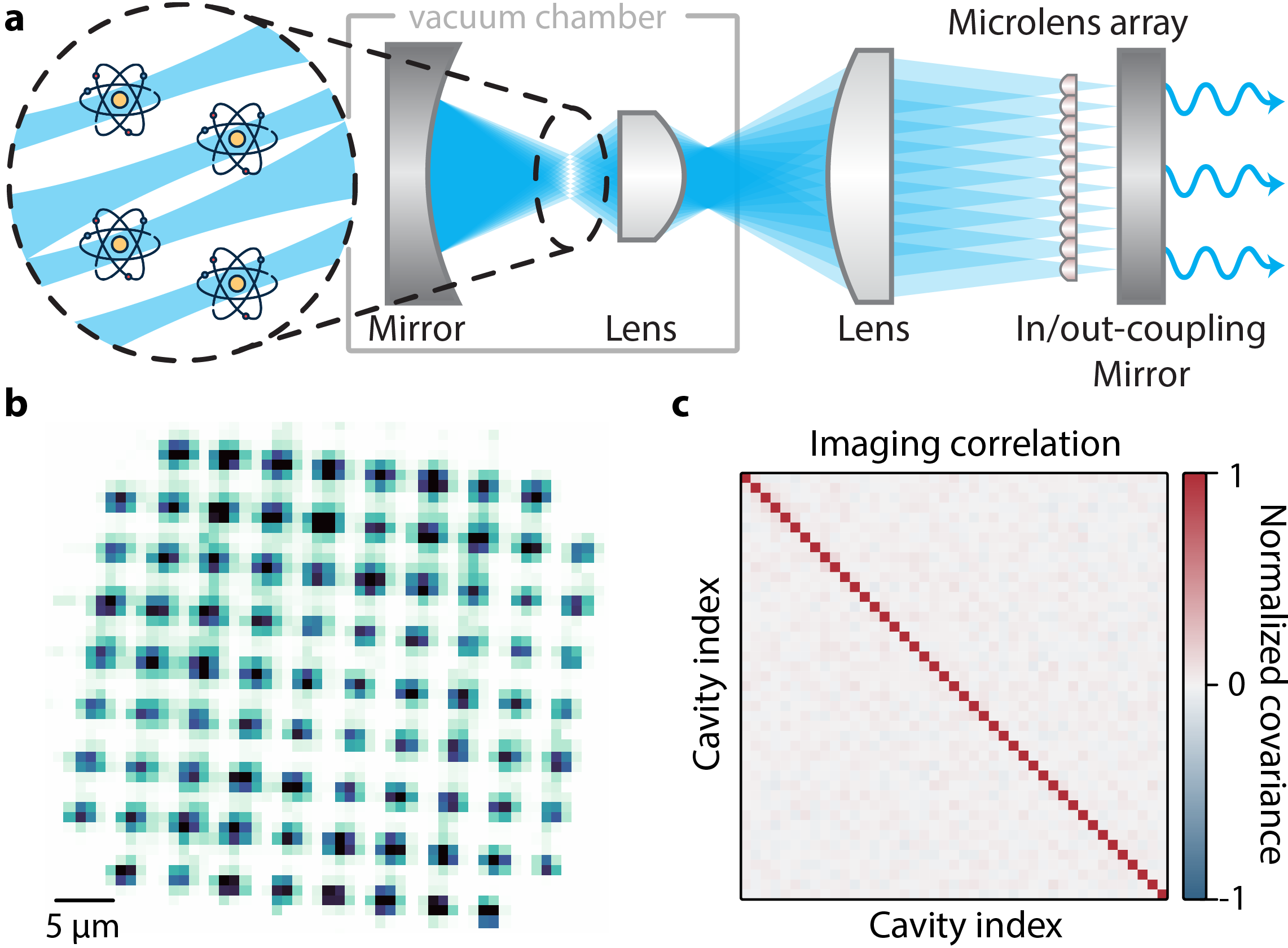}
	\caption{
	\textbf{The cavity array microscope.} \textbf{a.} We introduce a new type of cavity architecture, the cavity array microscope, which transduces between single atoms and single photons in parallel across a two-dimensional array of atoms. Our design uses intra-cavity lenses to engineer micron-scale cavity mode waists at the atom location, and an intra-cavity microlens array to engineer an array of such modes with micron-scale spacing, compatible with typical atom array geometries. \textbf{b.} Average atomic fluorescence image read out via each individual cavity mode. \textbf{c.} Cross-cavity correlations between measured photons are on average ${\lesssim}1\%$, indicating each cavity-atom pair is independent.
	}
    \vspace{2mm}
	\label{fig:schematic}
\end{figure} 

\begin{figure*}[t!]
	\centering
 	\includegraphics[width=172mm]{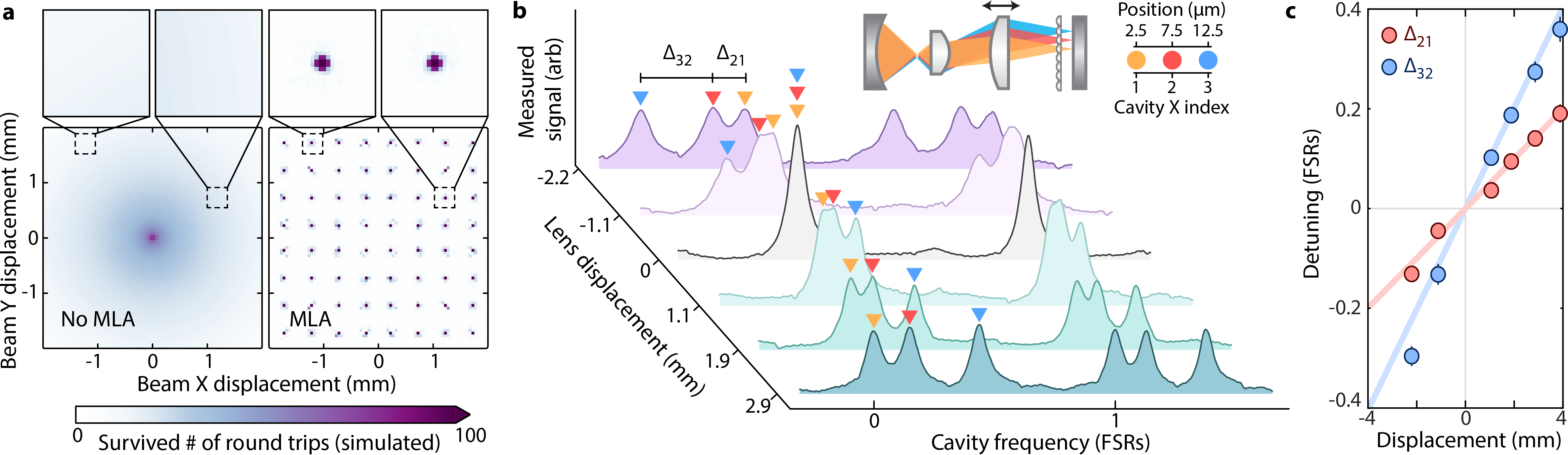}
	\caption{
	\textbf{Achieving a degenerate array of cavities.} \textbf{a.} Beams input to the cavity with some displacement from the central axis are not, by default, stably trapped due to intrinsic aberrations coming from the intra-cavity optics. This is evidenced by ray optics simulations (with lossless optics) of the number of round-trips an incident ray with zero slope makes within the cavity before escaping or clipping. However, adding an intra-cavity microlens array (MLA) provides local regions of stability even for large positional displacements. \textbf{b.} Inverted reflectance spectra over a free spectral range (FSR) of measured resonances for driving the fundamental Gaussian mode for three adjacent cavities (yellow, red, and blue; inset shows a partial round-trip for each mode) as the position of the intra-cavity spherical lens is tuned. For an arbitrary lens position the modes are separated, but by tuning to the position that creates a perfect telescope all modes are made simultaneously degenerate. The physical variable being swept on the $x$-axis is the voltage of the piezo-driven mirror, see Methods. \textbf{c.} Detunings (defined in \textbf{b}) between the Gaussian modes of adjacent cavities grow linearly as a function of lens displacement, with the sensitivity increasing as a function of the cavity's distance from the central axis.
	}
	\label{fig:novelty}
\end{figure*} 

In recent years, several experiments have begun interfacing atom arrays and cavities in order to realize these advantages~\cite{deist2022mid,grinkemeyer2025errordetected,hu2025siteselective,hartung2024quantumnetwork,liu2023realization,dhordjevic2021entanglement}, but in all cases have jointly coupled the entire atom array to one shared cavity mode. Reading out the state of an individual atom in this manner necessitates serialization with a time-cost scaling extensively with the system size, either through localized qubit addressing via ancillary lasers~\cite{hu2025siteselective,hartung2024quantumnetwork} or by physically moving atoms in and out of the shared cavity mode~\cite{deist2022mid}. Developing a truly parallelized atom-cavity coupled architecture is expected to not only improve networking rates~\cite{sinclair2025faulttolerant} and readout times~\cite{wang2025ultrafast}, but also has the potential to enable studies of new synthetic photonic materials~\cite{makin2008quantum,hartmann2008quantum,greentree2006quantum}. Multi-mode cavity designs have been explored in this context, but at the cost of single-atom resolution~\cite{vaidya2018tunablerange,leonard2017supersolid,clark2020observation}.

In this work, we present a new experimental platform which addresses this challenge, realizing a two-dimensional array of over 40 separate Gaussian cavity modes, each of which is coupled to a single atom (Fig.~\ref{fig:schematic}a). Each atom-cavity pair may be read out individually, which we demonstrate both on a traditional camera with high fidelity in millisecond timescales (Fig.~\ref{fig:schematic}b), as well as through a parallelized fiber array as a proof-of-principle for future networking applications~\cite{li2025parallelized}. Importantly, correlations between photon counts across the array are ${\lesssim}1\%$, indicating each atom-cavity pair is independent (Fig.~\ref{fig:schematic}c, Methods~\ref{met:datastatistics}).

The cavity is macroscopic (spanning ${\approx}34$ centimeters), employs exclusively off-the-shelf optics, and is primarily out of vacuum. Further, atoms are trapped at millimeter-scale distances from any dielectric surfaces, minimizing deleterious effects of surface charges for Rydberg-mediated interactions~\cite{ocola2024control}. These design features make it straightforward to integrate with standard atom array experiments---including with dynamic traps~\cite{endres2016atom,barredo2016atom}---in the future. In a glimpse ahead, we showcase (without atoms) a next-generation version of the system which demonstrates vast scalability in both system size and quality, achieving an array of over 500 cavities with a mean finesse greater than 100, while additionally being compatible with glass-cell-based experiments~\cite{soper2025cavity}. 

In total, our work develops new tools in cavity engineering to realize the power of parallel strong coupling of arrays of individual quantum emitters to arrays of optical cavities. Through these advances, our work signals a departure from traditional single-cavity experiments, and heralds a new era of many-cavity quantum information science. 

\subsection*{The cavity array microscope}
\label{sec:theplatform}
Our experimental platform consists of an array of trapped $^{87}$Rb atoms interfacing with the cavity array microscope architecture, shown in Fig.~\ref{fig:schematic}a (for detailed schematics, see Ext. Data Fig.~\ref{efig:detailed}; for summary of physical cavity parameters, see Ext. Data Table.~\ref{etab:parameters}; for details on atom loading and trapping, see Methods~\ref{met:atomloading}). We begin by describing the cavity optics and mode characteristics, before discussing the benefits for parallel atom readout below. We note that many aspects and metrics of the design are updated and improved in the next generation version of the system, presented toward the outlook~\cite{soper2025cavity}.

To illustrate how a single optical system realizes an array of independent, tightly spaced, micron-scale cavity modes, we first describe how trapping light travels through the system: an out-of-cavity spatial light modulator (SLM) generates an array of Gaussian beams with $100\ \upmu$m waists and $500\ \upmu$m pitch, focused onto the back of a flat cavity incoupling mirror. A two-lens 4f telescope inside the cavity, shared by all input beams, demagnifies this array by a factor of $M=100\times$ at the atom plane, where each beam focuses close to wavelength-scale~\cite{shadmany2025cavity}. The telescope is comprised of an out-of-vacuum spherical lens and an in-vacuum aspheric lens. Beams then reflect off of an in-vacuum curved mirror, spatially invert around the center of the array, and then return to the original flat incoupling mirror. After reflecting and re-tracing their paths in reverse, each beam has traversed a closed trajectory constituting a single local mode of the cavity.

The well-defined beam trajectories described above are only stable in an idealized system; real lenses are imperfect, inducing intrinsic aberrations and bringing with them technical imperfections like surface roughness that together lead to random diffusion of beam trajectories over many round-trips. This degrades cavity performance and eventually leads to clipping and loss. To stabilize the trajectories against such effects we introduce a microlens array (MLA), composed of a 20$\times$20 square grid of lenslets with $500\ \upmu$m pitch, in the image plane of the intra-cavity telescope. This optic breaks the spatial translational symmetry of the cavity, and provides local transverse confinement for the light~\cite{sommer2016engineering}, such that aberrations do not build up over many cavity round-trips.

To build intuition about this unique resonator-array geometry, we perform ray tracing simulations of the cavity array stability as a function of the location of an incident ray (Fig.~\ref{fig:novelty}a). Rays are allowed to propagate for up to 100 round-trips of the cavity, while we track if and when they are lost (due to clipping) during their transit. For input beams displaced from the central axis of the telescope we find that the number of round-trips falls off with radial distance; introducing the MLA produces regions of stability around each microlens even at large displacements from the central axis. Thus, the MLA allows us to maintain cavity performance out to large radial displacements, albeit restricting the cavity mode locations to a fixed geometry --- full ray tracing trajectories with the MLA are shown in Ext. Data Fig.~\ref{efig:detailed}. MLAs can be produced with many thousands of lenslets in customizable configurations~\cite{cai2021microlenses}, and since the optic is out-of-vacuum it can be easily swapped depending on the desired use-case. Non-uniformities between MLA lenslets can be actively compensated for using the SLM.

Importantly, all local Gaussian cavity modes can be made simultaneously resonant by adjusting the positions of the intra-cavity lenses into a 4f configuration (between the MLA plane and the atom plane) which minimizes aberrations. To find the optimal alignment, we use the SLM to initialize a row of three input beams coupled to adjacent fundamental Gaussian cavity modes. We then scan the longitudinal position of the intra-cavity, out-of-vacuum spherical lens, and record the resultant reflection spectra (Fig.~\ref{fig:novelty}b), from which we observe a clear optimum where all resonances completely overlap; spectra are obtained by dithering the location of the out-of-vacuum piezo mirror (Methods~\ref{met:cavitystabilization}). We find that the frequency deviations are linearly sensitive to spherical lens displacement (Fig.~\ref{fig:novelty}c), where the slope scales quadratically with the distance of the cavity mode from the telescope axis (Ext. Data Fig.~\ref{efig:sensitivity}). Reaching the degenerate condition requires no site-resolved control or feedback, and we predict that with micron precision in lens positioning, thousands of modes could be made simultaneously degenerate (Methods~\ref{met:degeneracy}). 

We emphasize that here we consider degeneracy between local Gaussian cavity modes, \textit{not} degeneracy of higher-order transverse modes for each local cavity; in fact, we observe that the higher-order transverse modes are far off-resonant with the local fundamental Gaussian mode of each cavity (Ext. Data Fig.~\ref{efig:abcdcalc}). Thus, the cavity array microscope is not a degenerate optical cavity in the conventional sense of degenerate transverse modes, such as in Refs.~\cite{vaidya2018tunablerange,clark2020observation}.

\begin{figure}[t!]
	\centering
 	\includegraphics[width=89mm]{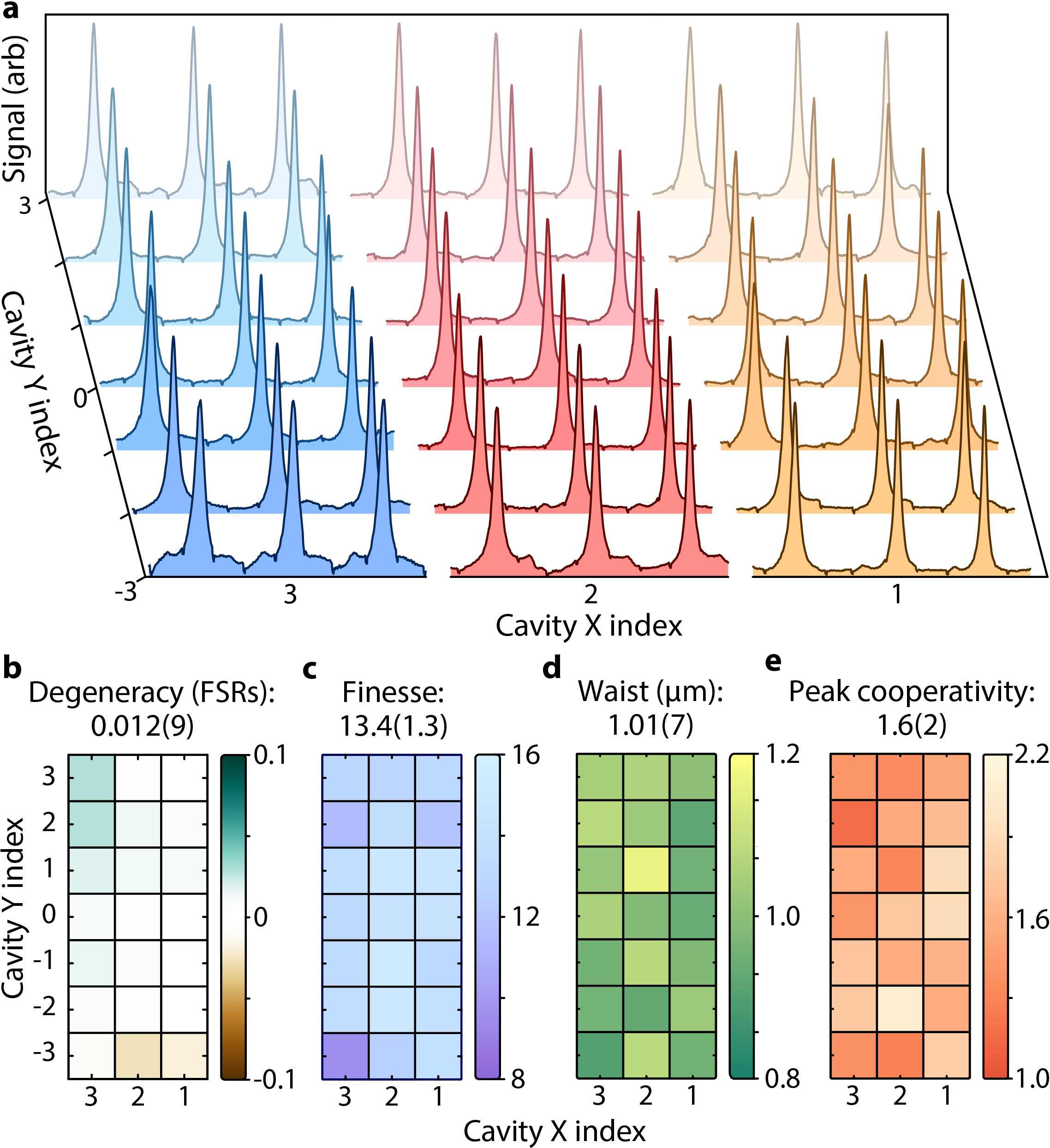}
	\caption{
	\textbf{Cavity array microscope performance.} \textbf{a.} Spectra showing three consecutive resonances for the central 21 cavity modes (measured sequentially), where the physical variable being swept is the voltage of the piezo-driven mirror. \textbf{b.} Degeneracy of the cavity modes. \textbf{c.} Finesses fit from cavity spectra. \textbf{d.} Cavity mode waists at the atom location, measured via trap depth and trap frequency spectroscopy. \textbf{e.} Calculated peak cooperativity. In \textbf{b-e} the array-averaged value (in \textbf{b}, absolute value) and standard deviation are under each subtitle.
	}
	\label{fig:cooperativity}
\end{figure} 

While we have described the system for the trapping and cavity-locking light, in our case at $\lambda=785$ nm, the same cavity optics support nearly identical modes at $\lambda=780$ nm~\cite{shadmany2025cavity}, the wavelength that is strongly coupled to our $^{87}$Rb atoms. We intentionally choose to realize both these applications---enhancement of the trap intensity and the imaging collection efficiency---in order to demonstrate both possibilities, but in general it is desirable to operate with larger trap detunings to reduce deleterious effects of non-scalar light shifts and incoherent scattering. With appropriate aberration correction, we imagine the cavity array microscope could be made more widely achromatic, but the system could also be designed to fulfill only one of these purposes. For instance, the cavity array microscope could be tuned for the atomic fluorescence transition while the traps are generated with standard optical tweezers injected via an intra-cavity dichroic or properly coated end-mirror, or vice-versa, thus preserving the desired benefits of the cavity array microscope for imaging, networking, or large-scale trapping.

The key figure of merit determining cavity performance is the cooperativity, which sets both the light collection efficiency $\chi=C/(1+C)$, and the number of coherent exchanges of information between atom and cavity field $N\approx \sqrt{C}/\pi$. For an atom located at the antinode of a standing wave cavity scattering on a closed transition, the peak (or geometric) cooperativity is given by~\cite{tanji2011interaction}
\begin{align}
C=6 F/\pi^3\times(\lambda/w)^2\ ,
\label{eq:cooperativity}
\end{align}
where $F$ is the cavity finesse, characterizing the number of round-trips the light makes through the cavity; and $w$ is the waist of the cavity mode at the atom location.

To measure the finesse, we use the SLM to individually excite each of the central 21 cavity modes in series, and obtain reflectance spectra  (Fig.~\ref{fig:cooperativity}a), finding all cavities are mutually degenerate to around a percent of the free spectral range (FSR), much smaller than the linewidth (Fig.~\ref{fig:cooperativity}b, Methods~\ref{met:description}). We fit the resultant cavity linewidth and FSR, from which we find an array-averaged finesse of $F=13.4(1.3)$, where the error bar is the standard deviation over cavities (Fig.~\ref{fig:cooperativity}c). This value is primarily limited by internal losses from various commercial intra-cavity optics and by in-vacuum misalignment incurred during the gluing and bake-out process (Ext. Data Fig.~\ref{efig:detailed}, Ext. Data Table~\ref{etab:losses}). 

To determine the cavity mode waist, we leverage the fact that in our present work the atom-traps are themselves cavity modes. To extract the trap sizes, we obtain the trap depth and trap frequency measured directly from the atoms (Ext. Data Fig.~\ref{efig:depth}). Assuming the traps are nearly harmonic, combining these two measurements fully determines the transverse mode waist~\cite{grimm2000optical}, from which we find an array-averaged waist and standard deviation of $w=1.01(7)\ \upmu$m (Fig.~\ref{fig:cooperativity}d), in good agreement with the predicted Gaussian mode waist of \mbox{$w_0=\frac{1}{M}\times \sqrt{\frac{f_{\textrm{MLA}}\lambda}{\pi}}\approx 1.08\ \upmu$m} obtained through ABCD matrix calculations, where $f_{\textrm{MLA}}=46.7\ $mm is the focal length of the MLA (Ext. Data Fig.~\ref{efig:abcdcalc}).

From these measurements of finesse and waist, we estimate the array-averaged peak cooperativity and standard deviation of $C=1.6(2)$ (Fig.~\ref{fig:cooperativity}e), already above unity thanks to the small mode waists across the array, with substantial headroom for improvement by increasing the finesse, demonstrated below.

\subsection*{Parallel readout}
\label{sec:readout}
To demonstrate the benefit of parallelized strong light-matter coupling afforded by the cavity array microscope, we exploit our ability to perform fast, non-destructive readout of individual atoms across the array simultaneously. We utilize $^{87}$Rb atoms, but we emphasize that the cavity array microscope is agnostic to choice of atom species. Fluorescence is driven by a pair of optical molasses beams that illuminate all atoms (Methods~\ref{met:atomloading}, Ext. Data Fig.~\ref{efig:2atom}). For these experiments the reflectivity of the cavity outcoupling-mirror is optimized for readout fidelity~\cite{shadmany2025cavity}, ensuring that the cavity-scattered photons leak out through this mirror, and can be directed onto an EMCCD camera.

\begin{figure}[t!] 
	\centering
 	\includegraphics[width=89mm]{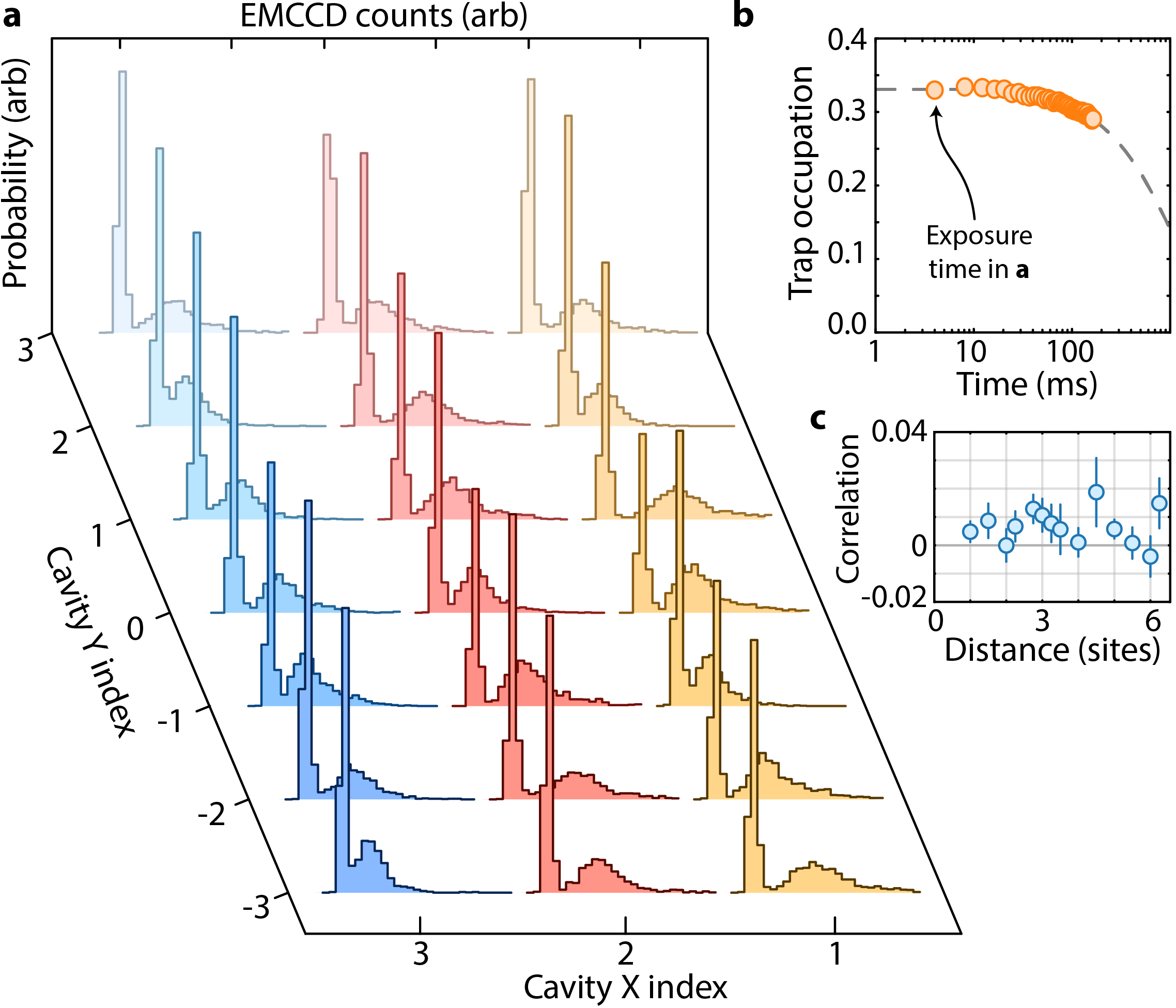}
	\caption{
	\textbf{Imaging with the cavity array microscope.} \textbf{a.} EMCCD fluorescence histograms (measured in parallel) with $4$~ms of exposure time, showing well-resolved bimodal distributions indicative of single-atom imaging. After post-processing~\cite{bergschneider2018spinresolved}, we find an array-averaged discrimination fidelity of $0.992(2)$. \textbf{b.} Trap occupation as a function of time during imaging, averaged across four atoms measured using fiberized readout (details in Fig.~\ref{fig:fiberarray}); fitted exponential decay (dashed line) yields an atom survival probability at $4$~ms of ${>}0.996$, limited by vacuum lifetime. \textbf{c.} Correlations in photon counts across the array are independent of distance and are uniformly ${\lesssim}1\%$.
	}
	\label{fig:imaging}
\end{figure} 

\begin{figure*}[t!]
	\centering
 	\includegraphics[width=178mm]{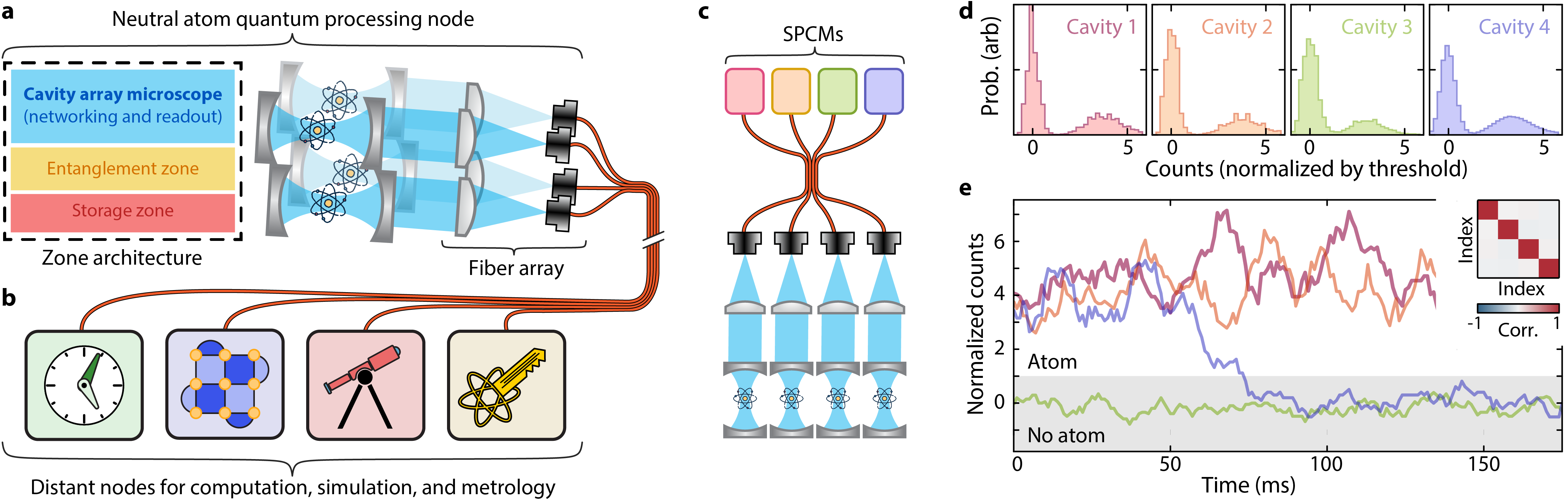}
	\caption{
	\textbf{Parallelized fiber readout for quantum networking.} \textbf{a.} We anticipate integrating the cavity array microscope into zone-based neutral atom quantum processors, using atom rearrangement to move atoms into and out of the cavity modes as a means of achieving fast readout and quantum networking. The cavity array microscope can be, in parallel, coupled to a fiber array such that each cavity mode leaks into a unique fiber. \textbf{b.} Networking photons between different nodes would enable many applications throughout quantum information science. \textbf{c.} As a first step towards this goal, we couple a four-mode cavity array to a four-channel fiber array, allowing individual fiber readout of each site using single photon counting modules (SPCMs). \textbf{d.} Imaging histograms; due to inhomogeneous dark rates in available SPCMs, imaging histograms are shifted by their `no-atom' baseline and normalized by their discrimination threshold. \textbf{e.} In the future, parallel fiber readout will be critical for entanglement distribution, and already can be used for making finely time-resolved measurements of trap occupations to resolve individual atom loss events. Inset: imaging through the fiber still shows $\lesssim1\%$ photon correlations for all sites. 
	}
	\label{fig:fiberarray}
\end{figure*} 

With a 4~ms exposure, we find well-resolved bimodal distributions of counts across the central 21 cavity modes (Fig.~\ref{fig:imaging}a). Post-processing to reduce readout noise~\cite{bergschneider2018spinresolved}, produces an array-averaged discrimination fidelity of $0.992(2)$ (Ext. Data Fig.~\ref{efig:processed}). From a continuous time-series measurement of atom population obtained with a single-photon-counting-module (SPCM) -- discussed in more detail below -- we estimate the atom survival in this time-frame is ${>}0.996$ (Fig.~\ref{fig:imaging}b), limited by the vacuum lifetime of a few seconds. For these images we study the distance-dependent correlations in the measured photons, which we find to be uniformly $\lesssim1\%$, indicating that the cavity modes are well isolated from each other (Fig.~\ref{fig:imaging}c, Methods~\ref{met:datastatistics}); these correlations could either arise from small (otherwise unobserved) cavity couplings, or, more likely, from residual common-mode environmental fluctuations. 

In this first demonstration, spatial inversion of the cavity mode due to reflection off of the curved mirror causes a single cavity mode to create two spatially separate traps in the atom and image planes. Thus, the 43 cavity modes in Fig.~\ref{fig:schematic}b produce 86 fluorescence spots. We do not explicitly prevent the two-atom loading, but work with intentionally low array filling to quadratically suppress two atom events, see Methods~\ref{met:atomloading}.

While the 4 ms of exposure time we use to achieve these fidelities is already faster than most typical non-destructive imaging in atom array experiments, it is far from the fundamental limit of the platform. Imaging parameters were optimized to ensure high survival in the long-imaging limit, not specifically for ultra-fast imaging. With readily achievable technical improvements to optics and imaging parameters, we expect $\approx10\times$ improved imaging times; see Methods~\ref{met:photonbudget} for a discussion of current and expected imaging performance. Substantial further gains are expected in the next-generation of the apparatus, showcased below. 

\subsection*{Fiber coupling and future networking}
\label{sec:fiberized}
The cavity array microscope enables more than just fast readout; high efficiency photon collection can also be used as the prototypical building block of a distributed quantum network~\cite{huie2021multiplexed,li2024highrate,covey2023quantum,zhang2025proposal,sinclair2025faulttolerant,hahn2025deterministic}. We imagine integrating the cavity array microscope as a constituent of zone-based neutral atom quantum processors~\cite{bluvstein2024logical}, using the next-generation design detailed in the following section to eliminate the in-vacuum curved mirror. With appropriate choice of mirror coatings or use of intra-cavity dichroics, dynamic optical tweezers could be generated and rearranged~\cite{endres2016atom,barredo2016atom} to move atoms into and out of the cavity readout modes with nanometer precision~\cite{deist2022superresolution,shaw2024multiensemble}. Further, the long working distance of our system ($\approx1$ mm for the aspheric lens) minimizes surface charge driven decoherence of Rydberg-mediated interactions~\cite{ocola2024control}, enabling readout nearer to entangling zones to minimize atom movement and improve processor rates. As each individual cavity is a single (Gaussian) mode (Ext. Data Fig.~\ref{efig:abcdcalc}), each can be coupled in parallel to a unique single-mode fiber in a fiber-array for photon distribution (Fig.~\ref{fig:fiberarray}a). Combined with appropriate means of cross-cavity photon interference, such a modality would enable a wide array of tasks throughout quantum information science (Fig.~\ref{fig:fiberarray}b). Importantly, the natural parallelization of our cavity array microscope is expected to dramatically improve networking rates~\cite{sinclair2025faulttolerant}.

As a first step towards this goal, we demonstrate coupling to, and readout from, a fiber array which is matched to the array of cavity modes (Fig.~\ref{fig:fiberarray}c). We employ a one-dimensional multimode fiber array with four individual cores laid out in a line, where each core in the array is directly connected to an SPCM. Similar fiber arrays are available with lensed single mode fibers~\cite{li2025parallelized} and in two-dimensional arrays with hundreds of individual cores. 

With a fiber coupling efficiency of ${\approx}65\%$ across the array we find similar imaging fidelities as were achieved with the EMCCD (Fig.~\ref{fig:fiberarray}d). With more careful mode matching and alignment, we expect substantially higher fiber coupling efficiencies are feasible as both the cavity mode and fiber mode are Gaussian, as opposed to fiberized photon collection directly through a microscope objective which can suffer from poor spatial overlap of the collected dipole radiation emission pattern and the fiber mode~\cite{li2025parallelized,stephenson2020highrate}. Imaging through the fiber array confers access to the fine-grained time dynamics of the atom signal, allowing us to simultaneously monitor the real-time occupation of traps across the array~\cite{peters2025cavityenabled} (Fig.~\ref{fig:fiberarray}e). Importantly, all cross-correlations are still $\lesssim 1\%$ (Fig.~\ref{fig:fiberarray}e inset, Methods~\ref{met:datastatistics}). 

\begin{figure}[t!]
	\centering
 	\includegraphics[width=87mm]{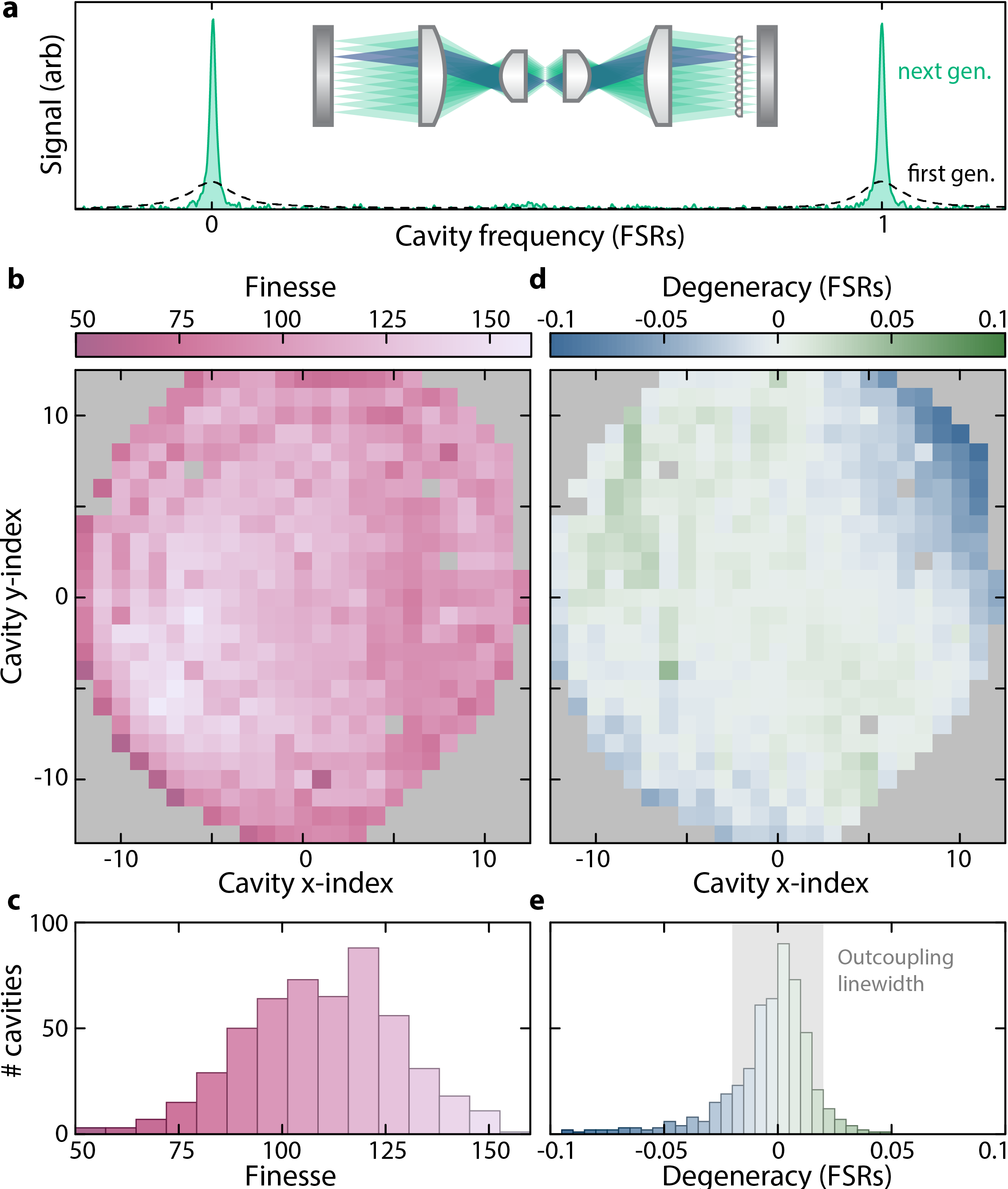}
	\caption{
	\textbf{A next-generation cavity array microscope.} \textbf{a.} Representative single cavity resonances of the next-generation and first-generation systems. Inset: the next-gen system replaces the curved mirror with a second 4f telescope, and uses improved optics; a single round-trip is highlighted for clarity. \textbf{b, c.} We achieve 516 resolvable cavities, with a mean (std. dev.) finesse of 110(18). \textbf{d, e.} Over 400 cavities are degenerate to within the optimized outcoupling linewidth (see text).}
	\label{fig:nextgen}
\end{figure} 

\subsection*{A next-generation design}
The presented cavity array microscope serves as a powerful prototype for the future of many-cavity QED studies, with immense potential to scale to far larger system sizes and finesses. As a glimpse towards this future, in an out-of-vacuum test setup we realize a next-generation variant of the system which achieves order-of-magnitude improvements along both axes (for further details, see Ref.~\cite{soper2025cavity}).

The system's design is similar, with a key change: we replace the curved end mirror with a second 4f telescope and flat end mirror (Fig.~\ref{fig:nextgen}a). This has multiple benefits, primarily that each cavity mode now has only a single small mode waist in the atom plane. All optics used in the cavity are improved, leading to substantially lower loss (Methods~\ref{met:nextgen}); the aspheric lenses also have an even longer working-distance of $\approx5.7$ mm, further reducing deleterious effects from surface charges when driving Rydberg excitations~\cite{ocola2024control}. In the future both aspheres could likely be replaced with out-of-vacuum microscope objectives typically employed in atom array experiments to enable full integration with glass-cell-based experiments.

We demonstrate the ability to individually resolve 516 cavities in the array with a mean (std. dev.) finesse of $110(18)$, an $>8\times$ improvement over the first generation design (Fig.~\ref{fig:nextgen}b). We believe the number of cavities is currently limited by the asphere field-of-view, which could be improved by using specially designed objectives in their place~\cite{manetsch2025tweezer}. Achieving degenerate operation is similar to in the first-generation cavity array microscope, but more stringently constrained by improved finesse and sensitivity to the transverse lens misalignment~\cite{soper2025cavity}. In order to optimize the cavity outcoupling, we would choose the outcoupling mirror to have a reflectivity of $82\%$ with an effective finesse of $\approx25$ (Methods~\ref{met:nextgen}). Over 400 cavities are degenerate to within the resultant linewidth (Fig.~\ref{fig:nextgen}c), primarily limited by minor astigmatism across the array that we believe originates from a mildly astigmatic mirror which we are now testing replacements for.

For these parameters, we estimate an optimized cavity collection efficiency of $55\%$, and a corresponding imaging time of $<100\ \upmu$s across the entire array simultaneously while maintaining high fidelity and atom survival (Methods~\ref{met:improvingphoton}). This performance is still far from what we envision as the limit of the platform -- for instance with further optimization of the cavity alignment the finesse could likely be made more homogeneous and overall higher across the array -- but we see it as an exciting next step into the frontier of many-cavity QED opened in this work.

\subsection*{Outlook}
\label{sec:outlook}
We have introduced the cavity array microscope, a new experimental platform which hybridizes atom arrays and optical cavities in a parallelized and scalable manner. Intrinsic to our designs is compatibility with existing atom array architectures and ease of integration with such systems. 

We expect this platform will be a key enabler of several important applications throughout quantum information science. By collecting photons via optical fibers, as we have demonstrated, cavity array microscopes could be efficiently connected across large distances as prototypical building blocks of a large-scale quantum network, or small distances as building blocks of a modular quantum computer. Within a single processing node the fast, parallel measurement afforded by cavity readout could greatly reduce the effective cycle time of neutral atom quantum processors close to nanosecond timescales~\cite{wang2025ultrafast}. Additionally, atom array experiments in general could benefit from trapping atoms directly into local cavity modes, as we do here, as a pathway to significantly improved scalability by taking advantage of intra-cavity power build-up to achieve many thousands of atom traps without requiring 100s of watts of laser power~\cite{manetsch2025tweezer}.

Beyond computing, the prospect of interconnecting multiple cavities, either locally or non-locally, could enable engineering of hybrid atom-photon Hamiltonians like the prototypical Jaynes-Cummings-Hubbard Hamiltonian~\cite{makin2008quantum,hartmann2008quantum,greentree2006quantum}, which describes a lattice of coupled cavities, each with a single respective two-level system, with photons itinerant across the array. Despite almost two decades of theoretical interest, this model has never been realized with optical photons. The cavity array microscope platform serves as a prime candidate for its realization in a large-scale, two-dimensional system -- all that is required is to intentionally induce cross-talk between adjacent cavity modes.

This work demonstrates the power of high-NA/low-finesse cavities for studying  cavity QED in the strong coupling regime~\cite{shadmany2025cavity}, where the vastly enhanced robustness to optical loss enables myriad new approaches via the introduction of intra-cavity optics. For instance, by introducing active elements into the cavity it may be possible to programmatically couple arbitrary sets of cavities; while such an active element would normally be nonviable in high finesse cavity QED experiments, in a cavity array microscope the loose requirement on finesse to achieve a high cooperativity is more forgiving of their integration.

Overall we expect that many practical applications will be realized by hybridizing cavity array microscopes with state-of-the-art neutral atom quantum processors, and that many new discoveries await in the burgeoning field of many-cavity QED which we study here.

\subsection*{Acknowledgments}
A.L.S. is supported by the Stanford Science Fellowship, and additionally by the Felix Bloch Fellowship and the Urbanek-Chodorow Fellowship. A.S. acknowledges support from the Hertz Foundation and the DoD NDSEG Fellowship. D.S. acknowledges support from the NSF GRFP. This work was supported by the National Science Foundation (NSF) through QLCI-HQAN grant 2016136, by AFOSR grant FA9550-22-1-0279, ARO grant W911NF-23-1-0053, AFOSR MURI Grant FA9550-19-1-0399, and AFOSR DURIP FA9550-19-1-0140. We acknowledge Henry Ando and Chuan Yin for work in construction of the vacuum system, and Hannah Manetsch for feedback on the manuscript.

\section{Author Contributions}

A.L.S., A.S., and D.S. performed the experiments and data analysis. A.L.S., A.S., D.S., A.K., L.P., and D.K. built the experiment. A.S., D.S., A.K., L.P., V.K, L.T., and M.J. contributed to prototypes of the experimental geometry. A.L.S. and J.S. wrote the manuscript with contributions and input from all authors. D.I.S. and J.S. supervised this project. A.L.S., A.S., and D.S. contributed equally.

\section{Competing Interests}
M.J. and J.S. act as consultants to, and hold stock options from Atom Computing. D.S., A.K., M.J., D.I.S. and J.S. hold a patent on the resonator geometry demonstrated in this work.

\clearpage



\bibliographystyle{adamref}
\bibliography{references.bib}


\subsection*{Data Availability}
The experimental data presented in this manuscript are available from the corresponding author upon request, due to the proprietary file formats employed in the data collection process.
\subsection*{Code Availability}
The source code for simulations throughout are available from the corresponding author upon request. Codes for cavity ray-tracing are available at Ref.~\cite{Palm_sloppy}.

\clearpage
\newpage





\renewcommand{\theequation}{S\arabic{equation}}
\renewcommand{\thefigure}{\arabic{figure}}
\renewcommand{\figurename}{Ext. Data Fig.}
\renewcommand{\figurename}{Ext. Data Fig.}
\renewcommand{\tablename}{Ext. Data Table}
\setcounter{figure}{0}
\setcounter{table}{0}
\setcounter{equation}{0}

\onecolumngrid
\newpage
\phantomsection
\section*{Extended Data Figures}
\FloatBarrier

\begin{figure*}[ht!]
	\centering
        \phantomsection
 	\includegraphics[width=175mm]{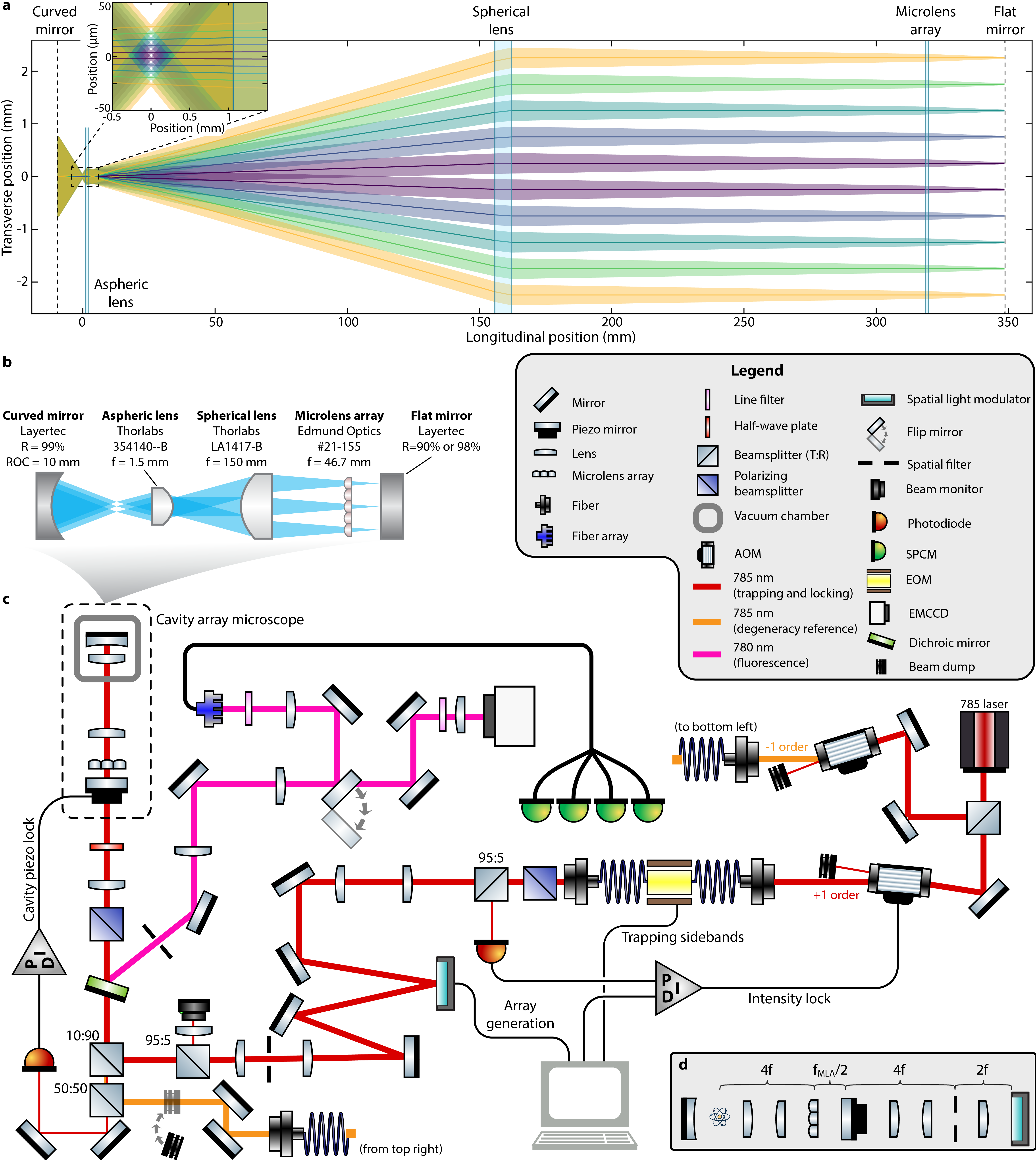}
	\caption{
    	\textbf{Detailed schematic of the cavity array microscope and surrounding experiment.} \textbf{a.} Full ray-tracing simulation of a one-dimensional slice of the cavity array microscope. \textbf{b.} Intra-cavity optics with manufacturer and part numbers.  For the flat end-mirror we either use a reflectivity of $98\%$ for measurements of the finesse (Fig.~\ref{fig:cooperativity}), or $90\%$ for measurements of atoms to increase the collection efficiency (Figs.~\ref{fig:imaging} and~\ref{fig:fiberarray}). \textbf{c.} Schematic of the experimental system surrounding the cavity. \textbf{d.} Simplified schematic of cavity array generation showing relevant telescopic distances.
	}
	\label{efig:detailed}
\end{figure*} 

\begin{table*}[t]
\centering
\begin{threeparttable}
    \flushright
    \caption{\textbf{Cavity array microscope parameters.} Values represent typical, array-averaged quantities, while parentheticals indicate the standard deviation across the array.}\label{etab:parameters}
    \vspace{0mm}
    \begin{tabular}{c l c c c c}
        \toprule
        \ \hspace{1mm} & \textbf{Mean parameter}\hspace{6mm} & \hspace{12mm}\textbf{Value} (first gen) & \hspace{2mm}\textbf{Value} (next gen) \hspace{12mm} & \textbf{Unit} &\  \hspace{1mm} \\
        \midrule
        &Length & 0.34\tnote{a} & 1.04 & m  & \\
        &Achieved \# of cavities & 43 & 516 & -- &\\
        &FSR             & 205 & 145 & MHz &\\
        &$\kappa$        & 15.3 & 1.4 & MHz &\\
        &$\Gamma$        & 6.1 & 6.1 & MHz &\\
        &$g$, peak\tnote{b}    & 6.0 & 4.4 & MHz &\\
        &Finesse, average (std. dev.)         & 13.4(1.3) & 110(18) & -- &\\
        &Waist, at focus & 1.01(7) & 1.15\tnote{c} & $\mu$m &\\
        &Pitch, at focus & 5 & 10 & $\mu$m &\\
        &Peak cooperativity & 1.6(2) & 9.8\tnote{c} & -- &\\
        &Peak collection efficiency & 18.1 & 55\tnote{c} & \% &\\
        
        \bottomrule
    \end{tabular}
    \begin{tablenotes}[para,flushleft]
    \footnotesize
    \raggedright
    \item[a] The physical cavity length is 34 cm, but the curved mirror inversion effectively doubles this length.\\
    \item[b] Inferred from the peak cooperativity.\\
    \item[c] Predicted.
\end{tablenotes}
\end{threeparttable}
\vspace{1cm}
\end{table*}

\begin{figure*}[ht!]
	\centering
        \phantomsection
 	\includegraphics[width=141mm]{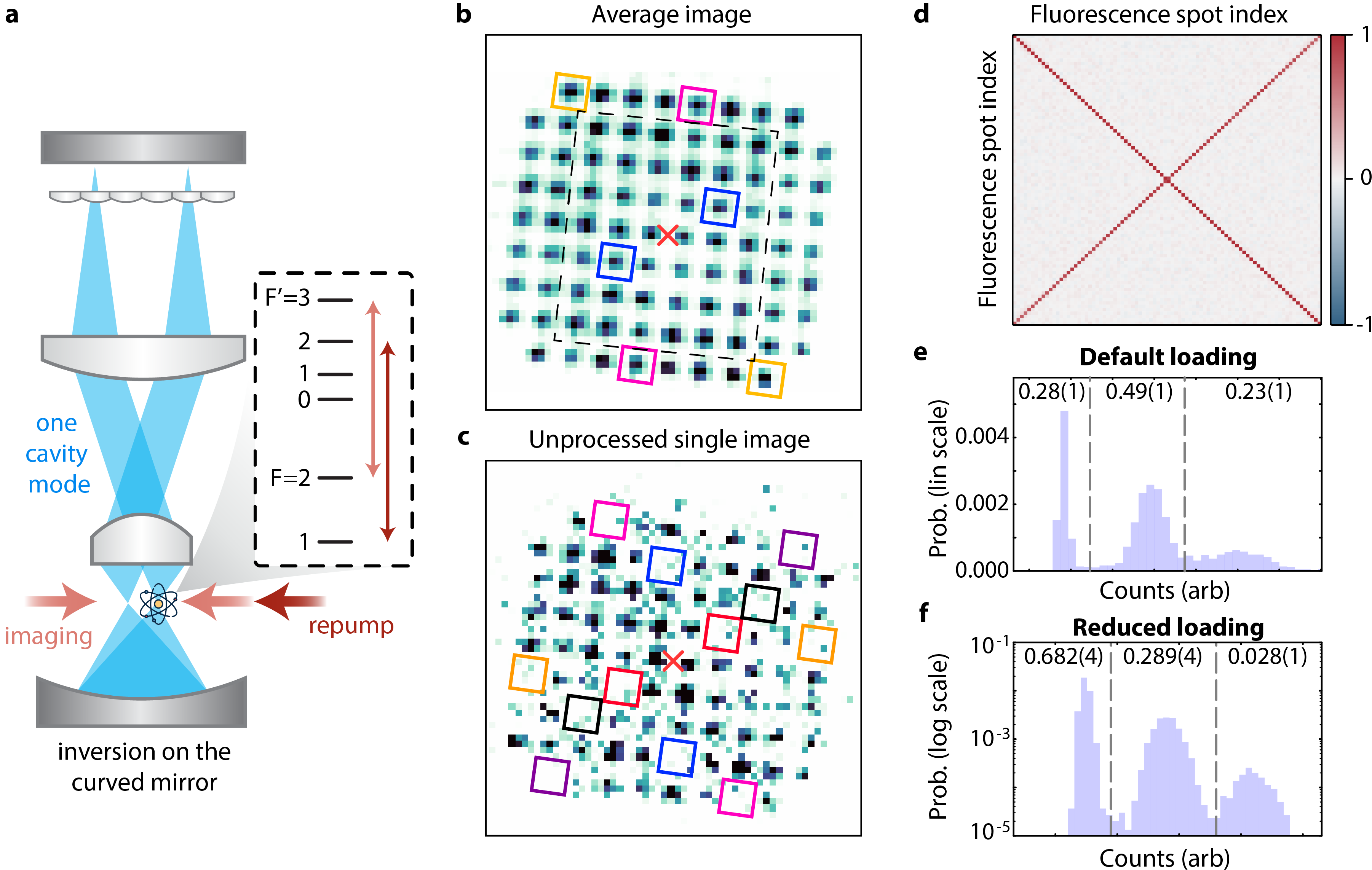}
	\caption{
    	\textbf{Doubled imaging and two-atom loading.} \textbf{a.} In the cavity geometry we introduce, inversion on the curved mirror leads to each cavity mode having two wavelength-scale waists, and two outcoupling ports. Also annotated are the directions and transitions of the beams driving fluorescence and atom repumping. \textbf{b.} The conjugate imaging ports are inverted around the center of the array (red cross), indicated here for a few select pairs of ports (colored boxes) on the average fluorescence image. The black dashed box indicates the central 21 cavities which we consider for most in-depth array characterizations in the main text. \textbf{c.} Single-shot fluorescence images show an inversion-symmetric pattern of fluorescence due to the doubling effect (conjugate ports for which no atom has been flagged are boxed with the same color to highlight the inversion). \textbf{d.} The doubling is clear from the near-perfect photon correlations between conjugate ports of fluorescence spots (ordering is column-major); in Fig.~\ref{fig:schematic}c of the main text, the correlation is taken after imaging counts have been summed over both output ports. \textbf{e.} By default, this doubling effect will also induce two-atom loading, one in each wavelength-scale waist, as evidenced by a tri-modal fluorescence histogram. 0-, 1-, and 2- atom loading fractions are annotated in each section. \textbf{f.} We intentionally lower the loading efficiency by lowering the trap depth during loading and pulsing traps on and off to quadratically suppress the double loading effect by a factor of $10\times$ compared to the single loading probability.
	}
	\label{efig:2atom}
\end{figure*}

\begin{figure*}[ht!]
	\centering
        \phantomsection
 	\includegraphics[width=93mm]{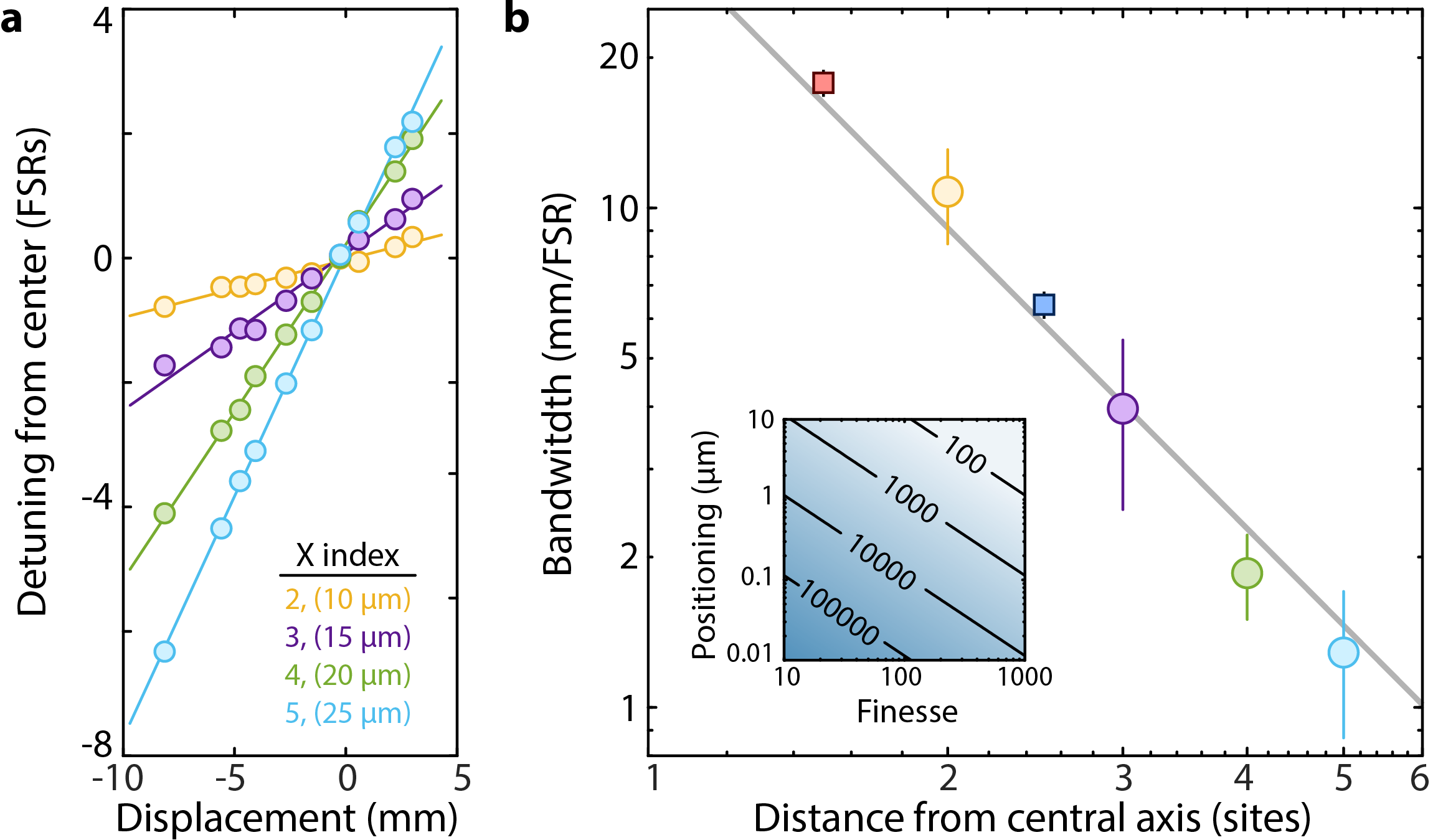}
	\caption{
	\textbf{Sensitivity of the lens degeneracy condition.} \textbf{a.} Detuning between the given cavity resonance vs the resonance of a cavity aligned to the telescope central axis (at $0\ \upmu$m). All cavities reach the same degeneracy condition, but with a sensitivity which increases with cavity $X$ index. \textbf{b.} The alignment bandwidth (the number of millimeters of lens displacement before the cavity mode moves by a full FSR, i.e. the inverse slope of the fits in \textbf{a}) decreases quadratically with distance from cavity center. Square markers are taken from Fig.~\ref{fig:novelty}c, where the data was taken relative to a cavity at $2.5\ \upmu$m. Inset) From Eq.~\eqref{eq:arraysize}, we estimate the maximum circular array size that could be made simultaneously degenerate as a function of the cavity finesse, $F$, and the minimum lens positioning accuracy; with micron-level positional accuracy, thousands of modes could be simultaneously degenerate at $F>100$. However, we emphasize that a more thorough study of how these values change as a function of lens type, focal length, MLA pitch, etc. is warranted, as the scaling performance may be significantly improved by using a reduced-distortion lens like an asphere or a low loss objective.
	}
	\label{efig:sensitivity}
\end{figure*} 

\begin{figure*}[ht!]
	\centering
        \phantomsection
 	\includegraphics[width=89mm]{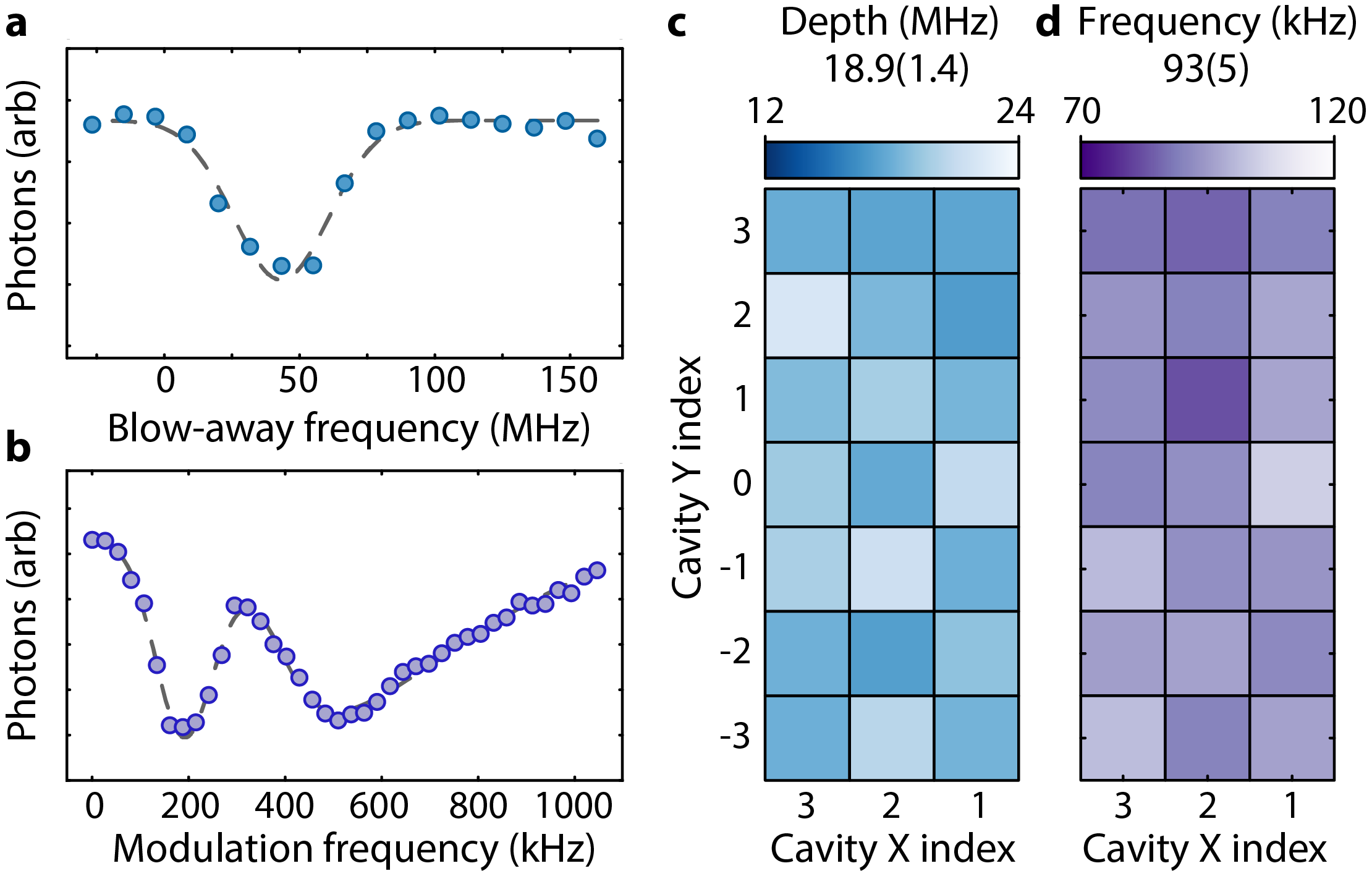}
	\caption{
	\textbf{Trap characterization.} \textbf{a.} Array-averaged trap depth measurement. Atoms are loaded, then strongly driven on the $F=2\rightarrow3$ transition to blast them away, then imaged at normal intensities. The differential light shift of the D2 transition is evident as the low photon feature. \textbf{b.} Array-averaged trap frequency measurement. Atoms are loaded, then their trap depth is modulated at varying frequency, leading to parametric heating at $2\times$ the transverse and longitudinal trapping frequencies. As the trapping potential is somewhat lattice-like the longitudinal trap frequency is higher than the radial trap frequency, and can be tuned by changing the intensity of sidebands applied to the trapping light~\cite{shadmany2025cavity}. \textbf{c,d,} Array-resolved trap depth and frequency; array-average and standard deviation is annotated under each subtitle.
	}
	\label{efig:depth}
\end{figure*} 

\begin{figure*}[ht!]
	\centering
        \phantomsection
 	\includegraphics[width=93mm]{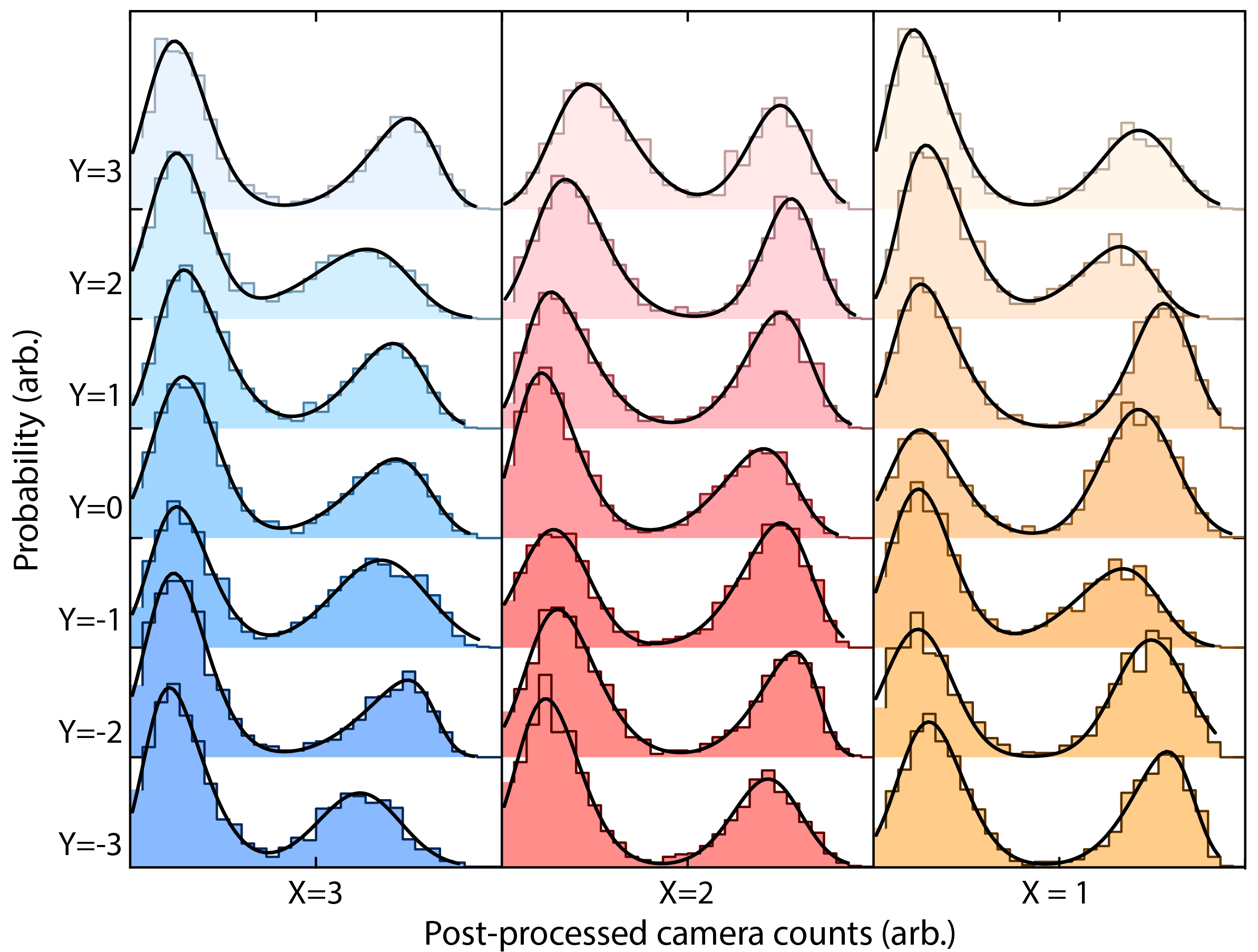}
	\caption{
	\textbf{EMCCD post-processed histograms.} To maximize fidelity, we post-process EMCCD images via thresholding and Gaussian smoothing methods~\cite{bergschneider2018spinresolved}. Fitting with a double-skew-Gaussian model (lines) yields an array-averaged discrimination fidelity of $0.992(2)$, while fitting with a less conservative double-Gaussian model yields $0.997(2)$.
	}
	\label{efig:processed}
\end{figure*} 

\begin{figure*}[ht!]
	\centering
        \phantomsection
 	\includegraphics[width=90mm]{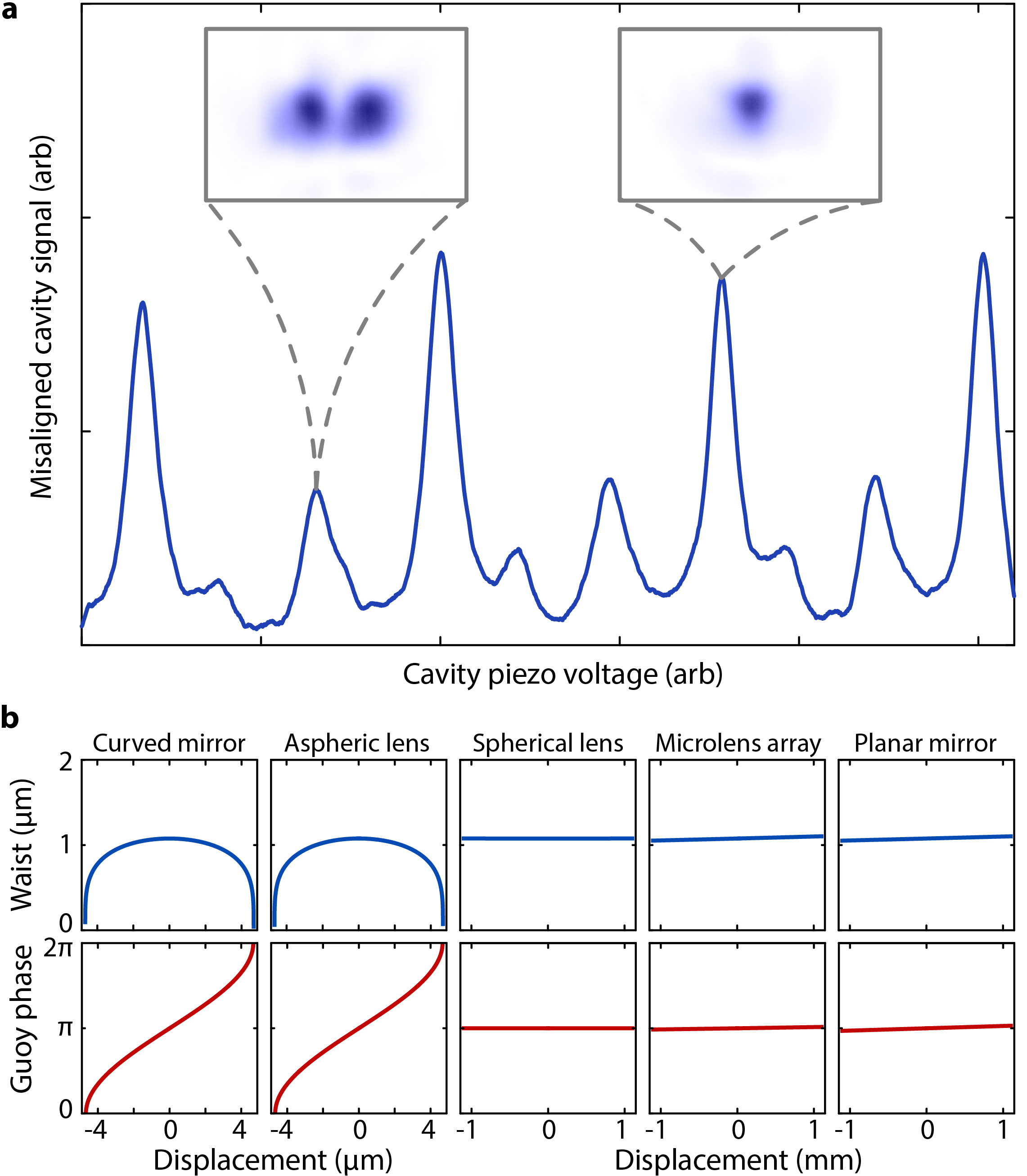}
	\caption{
	\textbf{Transverse mode splitting and cavity stability.} \textbf{a.} Transverse mode splitting for a single cavity, taken after significantly misaligning the incoupling of the laser driving the cavity to make higher-order modes more visible. The transverse mode splitting is controlled by the exact asphere-curved mirror separation, and for typical values the higher-orders are clearly off-resonant with the fundamental Gaussian mode (inset shows transmission images of the cavity modes). Higher-order modes are further suppressed by enhanced clipping losses. \textbf{b.} Mode waist and Gouy phase under longitudinal perturbations of the various optics in the cavity array microscope, calculated under a paraxial approximation. At the center of the asphere and curved mirror stability diagram, the waist is quadratically insensitive to displacements. For all other optics, the waist is insensitive over millimeter-scale displacements.
	}
	\label{efig:abcdcalc}
\end{figure*}

\begin{figure*}[ht!]
	\centering
        \phantomsection
 	\includegraphics[width=158mm]{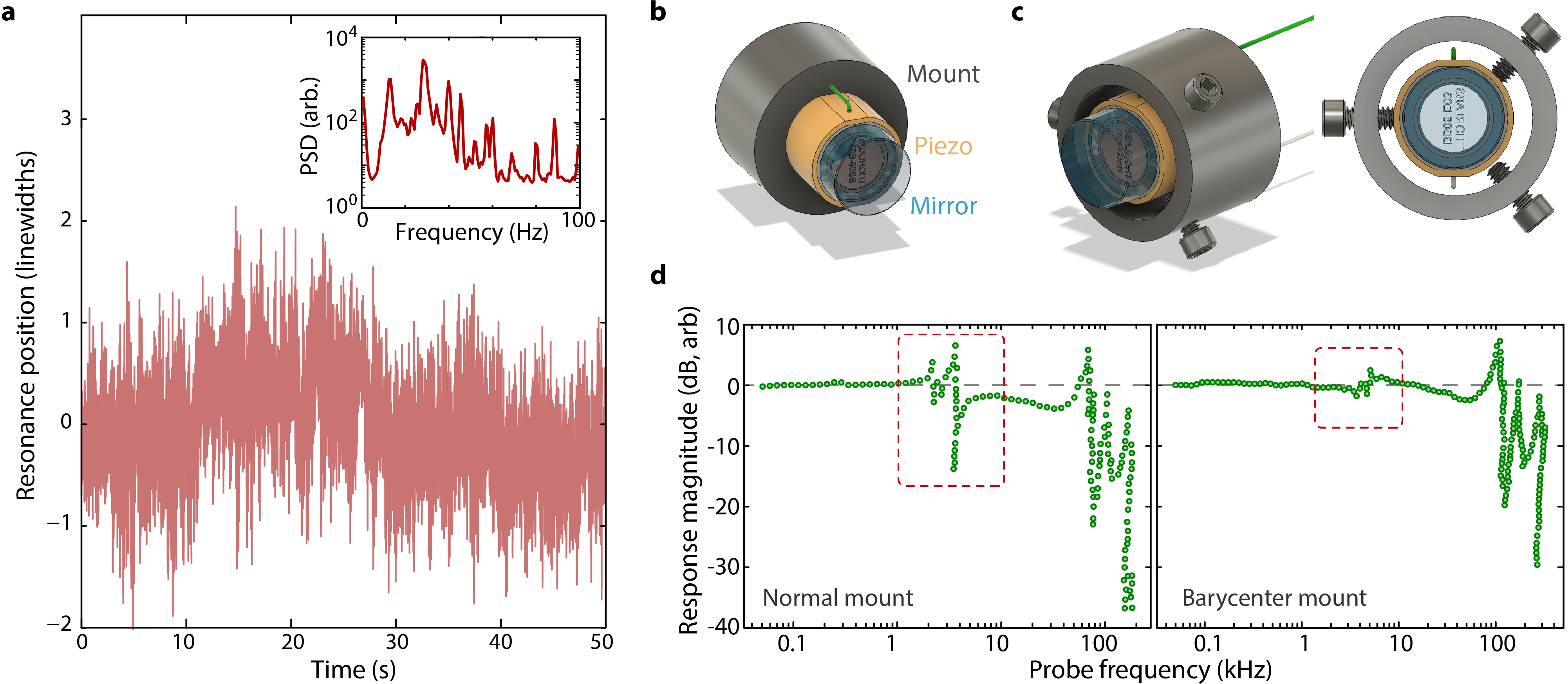}
	\caption{
	\textbf{Cavity locking with a barycenter mounted piezo.} \textbf{a.}  The in-vacuum curved mirror and asphere are jointly mounted on a meter long arm as part of the overall vacuum design~\cite{yin2023cavity}, which induces significant low frequency ``shaking'' of the frequency of cavity mode, as evidenced by the location of a single cavity resonance measured over tens of seconds (inset shows frequency power spectral density with strong components up to 100 Hz). \textbf{b.} We use the piezo-driven cavity end-mirror to lock the cavity length. Normal piezo mirrors are typically glued from the rear onto a mount, a configuration which means piezo expansion is unidirectional, increasing susceptibility to low-frequency noise. \textbf{c.} We adapt a scheme~\cite{chadi2013simple} where the piezo is clamped at its barycenter, such that expansion is bidirectional, reducing coupling to low frequency mount resonances. \textbf{d.} Interferometrically measured piezo-mirror frequency response function, showing the barycenter mount reduces the amplitude of low frequency mount resonances around a few kHz (highlighted with red dashed line), which allows the locking feedback loop to be operated at overall higher gain and noise suppression. We achieve a locking bandwidth of approximately 10 kHz, significantly suppressing the low frequency shaking noise. The remaining higher frequency resonances come from the self-resonance of the piezo itself.
	}
	\label{efig:barycenter}
\end{figure*} 

\begin{figure}[ht!]
	\centering
 	\includegraphics[width=86mm]{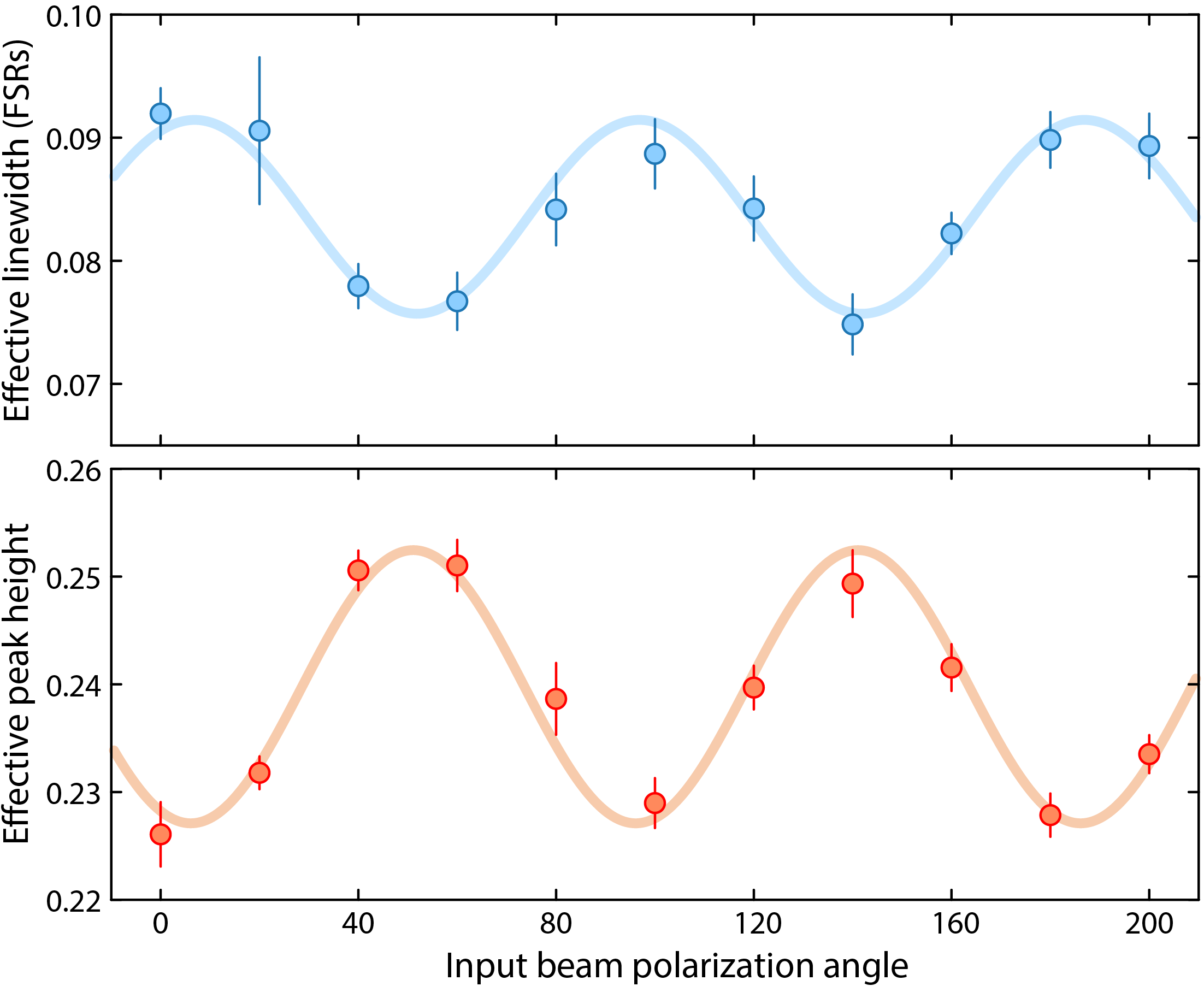}
	\caption{
	\textbf{Cavity birefringence.} We do not observe a resolvable splitting of the cavity resonance, but do observe subtle birefringent effects. In particular, varying the input beam polarization, we see the cavity linewidth and peak height show characteristic anti-correlated oscillations as one or both of the two polarization modes are alternatively driven. From the amplitude of the linewidth oscillations, we infer the unresolved splitting is $\approx 0.01$ FSRs.
    }
	\label{efig:bire}
\end{figure} 

\clearpage
\newpage
\twocolumngrid
\section*{Methods}
\appendix
\setcounter{secnumdepth}{3}

\makeatletter
\def\p@subsection{}%
\def\p@subsubsection{}%

\def\thesubsection{S\arabic{subsection}}%
\def\thesubsubsection{S\arabic{subsection}.\arabic{subsubsection}}%

\def\theHsubsection{S.\arabic{subsection}}%
\def\theHsubsubsection{S.\arabic{subsection}.\arabic{subsubsection}}%
\makeatother

\subsection{Description of the experiment}
\label{met:description}
Here we describe the fundamentals of the experimental system: the manner of cavity stabilization, the means of atom loading and imaging, and the generation of the cavity array. See Ext. Data Fig.~\ref{efig:detailed} for a schematic and layout of the experimental system for reference.

\subsubsection{Cavity stabilization}
\label{met:cavitystabilization}
\textit{Cavity locking ---} The cavity array presented in this work can be locked using a similar scheme to conventional two-mirror cavities with a piezo-electric motor on one end mirror to stabilize the overall length to vibrations and other noise sources. In particular, we pick off a small percentage of the tweezer light at 785 nm after it reflects off the cavity incoupling mirror, and direct it onto a photodiode from which we can measure the cavity spectra for all cavities simultaneously, see Ext. Data Fig.~\ref{efig:detailed}. We use a traditional Pound-Drever-Hall locking scheme with relatively small 10 MHz modulation sidebands to generate our error signal. This signal is delivered to a home-made locking servo, which feeds back on the out-of-vacuum piezo end-mirror of the cavity, with an achieved locking bandwidth of around 10 kHz (see below for further details). Ensuring the system is in a properly degenerate condition is important, as described in Fig.~\ref{fig:novelty} of the main text, such that cavity resonances are overlapping, and so the cavity feedback will stabilize all simultaneously. 

\textit{Compensating for mechanical noise ---} 
The relatively long overall physical cavity length of 34 cm necessitates installing half of the cavity in-vacuum and half outside of vacuum. The in-vacuum half of the cavity sits on a meter-long translator stage~\cite{yin2023cavity}, which we speculate is susceptible to large low-frequency mechanical noise because of its length. We directly observe these effects in Ext. Data Fig.~\ref{efig:barycenter}, where we monitor the position of a single cavity resonance over several tens of seconds. We find the resonance location `shakes' with a forest of modes at a few tens of Hz frequency. This effect is common-mode for independent lasers addressing the cavity, further suggesting it is a mechanical effect in origin.

Stabilizing our cavity against this large low frequency noise presents a challenge, due to the limited locking bandwidth of traditional piezo mount structures. Typically, the piezo-electric motor is affixed to one side of a mirror mount with the cavity mirror attached on the other. The asymmetry of this configuration leads to pronounced low frequency mount resonances that effectively cap the overall gain and noise suppression that the lock can be stably operated at. To circumvent this effect, we engineer a so-called ``barycenter mount''~\cite{chadi2013simple}. We insert the piezo in a metal ring as shown in Ext. Data Fig.~\ref{efig:barycenter} and clamp from the side with screws at its center of mass. The stack is then free to expand equally in both directions, effectively canceling out the net motion that would otherwise couple the longitudinal piezo motion to the elongation modes of the mount. 

We test the efficacy of this technique by using a network analyzer to measure the output of a Mach-Zehnder interferometer, one arm of which includes the piezo mount. We then characterize the mount resonances as the piezo is clamped at the center of mass of the piezo-mirror system. As shown in Ext. Data Fig.~\ref{efig:barycenter}, we see the relevant mount resonances at ${\approx}5$ kHz disappear almost entirely with the barycenter mount, indicating the successful suppression of the elongation mount modes when clamped at the center of mass. This technique allows us to engineer a robust lock for our resonator, which we tune to a bandwidth of approximately 10 kHz so as to be substantially below the remaining piezo self-resonance at approximately 100 kHz. This allows us to almost entirely eliminate the low-frequency shaking noise.

\textit{Sweeping cavity resonances---} In order to obtain cavity reflectance spectra (e.g. in Fig.~\ref{fig:novelty}) we directly vary the cavity length by dithering the voltage of the piezo incoupling mirror while the system is not in lock. As the stroke of the piezo is a few micron, we are able to scan through a few FSRs of the cavity in this way, which are then recorded directly onto the same photodiode used for locking. When needed, we convert between voltage applied to the piezo and cavity FSR in frequency by applying fixed frequency sidebands to the cavity driving light via a fiber EOM, which are then visible on the cavity reflectance and serve as a frequency ruler.

The homemade high-voltage piezo driver we use is not perfectly linear, and over typical ranges we do observe some non-linear slewing of the piezo; however, this effect at most contributes to $\lesssim3\%$ error in measured finesses, and typically we operate in a narrow region of the driver's output to maintain a consistent response.

\textit{Measuring cavity degeneracy---} In Fig.~\ref{fig:cooperativity} of the main text, we measure the degeneracy of the central 21 cavity modes for the first generation cavity array microscope. To do so, we incouple a second drive laser to the central cavity using a beam with a frequency offset, and measure the detuning from the sequentially excited cavities to this reference. Subtracting off the reference offset, we find all cavities are mutually degenerate to around a percent of the free spectral range (FSR), much smaller than the linewidth (Fig.~\ref{fig:cooperativity}b). The degree of degeneracy is limited by hand positioning of the spherical lens position with ${\approx}100\ \upmu$m precision (Ext. Data Fig.~\ref{efig:sensitivity}).

\subsubsection{Atom loading and imaging}
\label{met:atomloading}
\textit{Effective dipole trap ---} Details on the vacuum system, magneto-optical trapping (MOT) and atom delivery system have been described previously~\cite{yin2023cavity,shadmany2025cavity}. In brief, we form a MOT of rubidium 87 atoms, and then use a transport lattice to extract a group of atoms approximately 15 cm away into the area of the cavity optics. Atoms are loaded directly from the transport lattice into the cavity mode. In order to induce parity projection to ensure only a single atom per diffraction limited waist of the cavity mode, we drive the cavity with multiple tones: a carrier frequency, and sidebands $\pm10$ FSRs detuned. These sidebands effectively smear out the cavity potential, making it look somewhat like a dipole trap~\cite{shadmany2025cavity}. While the effect is not perfect (for instance, we still find the longitudinal trapping frequency is greater than the radial trapping frequency, Ext. Data Fig.~\ref{efig:depth}), we find that it still enables single atom loading with high probability. 

To determine the transverse waist of the traps, we measure both the trap depth and trap frequency (Ext. Data Fig.~\ref{efig:depth}). For the depth, $U$, after atoms are loaded they are strongly driven on the $F=2\rightarrow3$ transition to eject them from the trap, then imaged as normal. The resulting imaging signal shows a loss feature given by the differential light shift of the ground and excited states; at our trapping wavelength, the trap depth of the ground state is $43\%$ of the overall light shift~\cite{steck2024rubidium}. For the frequency, $\nu$, after loading the traps are modulated at fixed amplitude with varying frequency to induce parametric heating at twice the trapping frequency. After fitting the resulting $U_i$ and $\nu_i$ values for all traps in the array, we then calculate the waist assuming an approximately harmonic potential as $w_i=\frac{1}{\pi\nu_i}\sqrt{\frac{U_i}{m}}$, where $m$ is the atom mass~\cite{grimm2000optical}.

\textit{Single-atom loading ---} 
\label{met:singleatom}
In the first generation cavity array microscope, spatial inversion of the cavity mode due to reflection off of the curved mirror creates a point symmetry which causes a single cavity mode to create two spatially separate traps in the atom plane, and two output ports on the flat end mirror (Ext. Data Fig.~\ref{efig:2atom}). This is why the 43 cavity modes in Fig.~\ref{fig:schematic}b produce 86 fluorescence spots. This effect is eliminated in the next-generation cavity geometry~\cite{soper2025cavity} which has no curved mirror which we showcase in Fig.~\ref{fig:nextgen} of the main text. For the first-generation design, we take a more brute-force approach by quadratically suppressing the rate of double loading. If the probability of loading a single atom into one side of the cavity is $p$, then the probability to load only one atom across both sides is $2p(1-p)$, while the double loading probability is $p^2$. Thus by setting $p\approx0.18$ we should maintain a single atom loading probability of $2p(1-p)\approx30\%$, with a double loading probability of only $3\%$. Indeed, when we intentionally lower the loading fraction of the array by using a lower trap depth during loading and by pulsing traps on and off to eject atoms, we find that the probability to load only a single atom in either of the two sides of the array is $28.9(4)\%$, and the corresponding double loading probability is $2.8(1)\%$ (Ext. Data Fig.~\ref{efig:2atom}f); we consider this $10\times$ suppression of double-loading sufficient for our present purposes.

The largest array size in the original-generation apparatus we generate has 43 modes (Fig.~\ref{fig:schematic} of the main text) with a $40\ \upmu\textrm{m} \times 45\ \upmu\textrm{m}$ footprint in the atom plane, but this is close to saturating the available field-of-view. Several optics together enforce this limitation, including the limited clear aperture of the incoupling mirror and Glan-Taylor polarizer, and the field-of-view of the off-the-shelf aspheric lens. For results in Figs.~\ref{fig:cooperativity} and~\ref{fig:imaging} we focus on a subset of 21 modes with a $25\ \upmu\textrm{m} \times 30\ \upmu\textrm{m}$ footprint to avoid deleterious effects from edge modes on the verge of clipping (Ext. Data Fig.~\ref{efig:2atom}). These limitations are in no way fundamental, as we show in the next-generation cavity where we achieve an order of magnitude more modes, and where we believe the size limit is dominated by the asphere field-of-view~\cite{soper2025cavity}.

\textit{Cooling and imaging ---} In order to cool and image atoms, we employ a pair of co-aligned, oppositely propagating beams transverse to the cavity axis in order to minimize errant scatter and backgrounds. Each beam is red detuned 40 MHz from the $F = 2 \rightarrow F' = 3$ transition. A single pass beam (co-aligned with the cooling beams) tuned on resonance with the $F = 1 \rightarrow F' = 2$ transition repumps atoms into the $F = 2$ manifold. See Ext. Data Fig.~\ref{efig:2atom}a for a schematic of the level structure and the imaging geometry. We compensate for magnetic field gradients with three out-of-vacuum bias coils to operate at 0 field. The same parameters are used for PGC/light assisted collisions to load single atoms as for fluorescence imaging.

For imaging through the fiber array, to account for heterogeneity in dark-rates of our SPCMs, counts are shifted by their mean `no-atom' value and normalized by the optimal discrimination threshold. Further, we use a 10 ms exposure to account for both the finite fiber coupling efficiency of ${\approx}65\%$ across the array, and the $50\%$ drop in collection because each fiber is only coupled to one output port of each cavity mode (Fig.~\ref{fig:fiberarray}d, Methods~\ref{met:singleatom}). The mode-field-diameter (MFD) of each fiber core is $50\ \upmu$m, with a spacing of $127\ \upmu$m, chosen to roughly match the $40\%$ MFD-to-spacing ratio of the cavity modes at the atom and image planes.

\subsubsection{Polarization effects}
\label{met:polarization}

\textit{Tweezer polarization ---} In our current experiment, atoms are trapped at 785 nm, only 5 nm detuned from the atomic line at 780 nm. This leads to large vector stark shifts which makes the atom quite sensitive to trapping polarization. Unfortunately the dichroic mirror we use for separating the 785 nm trapping light from the 780 nm fluorescence signal is a simple tilted 808 nm line filter, and is thus quite angle-sensitive at the tilt angle required for splitting the 785 from the 780. The dichroic is roughly in the Fourier plane of the atom array, and there are up to degrees difference in the incident slopes of different beams, leading to polarization gradients across the array, and corresponding inhomogeneity in atom trapping. 

We take a maximalist approach to alleviating this effect by placing a Glan-Taylor polarizing beamsplitter between the dichroic and the cavity array microscope: this ensures homogeneous polarization and trapping across the array, at the cost of discarding half of the imaging light, which we accept in this current iteration of the experiment. We emphasize this is in no way a fundamental limitation, and could be completely eliminated by a variety of approaches, some of which are: 1) using a custom dichroic better suited for separating 780 vs 785 without angle-induced birefringence, 2) using further detuned trapping light, 3) moving the dichroic into the image plane of the array such that all beams have the same slope, 4) collecting light from both ports of the beamsplitter. 

\textit{Cavity birefringence ---} We find that the first-generation cavity array does exhibit some birefringent effects. In particular, when we vary the polarization angle of the beam incoupling the central cavity, we see the effective linewidth and resonance height oscillate out-of-phase with each other (Ext. Data Fig.~\ref{efig:bire}). We interpret this oscillation as an alternation between which, or both, of the polarization modes is driven. From the amplitude of the oscillations of the effective linewidth, we infer the mode splitting is $\approx0.01$ FSRs, well below the first-generation cavity linewidth. 

Further investigations are warranted into what potential misalignment or material effects lead to this birefringence, as it may affect performance for polarization-based imaging or networking protocols, especially for higher finesse systems like our next-generation design. We note in practice, though, that the finesse of the next generation will be intentionally lowered by a few factors of unity from its maximum to optimize outcoupling, making small splittings less deleterious. 

If some residual birefringence is difficult to neutralize, a benefit of the cavity array microscope architecture is that we can imagine inserting an intra-cavity birefringent element with a phase delay and rotation engineered to entirely cancel the existing effect with minimal finesse loss. Exploring such options will be an interesting and important area of future study.

\subsubsection{Array generation}
\label{met:arraygeneration}
We use a spatial light modulator (SLM, Meadowlark Optics) in order to generate the array of incoupling beams which drive the cavity array modes. While this is a standard technique in atom arrays, driving the cavity array becomes complicated because of the strict requirement of precision positioning of the input beams to properly incouple the cavity mode. To accomplish this, we directly generate each array spot at a precise position following the techniques in Ref.~\cite{chew2024ultraprecise} by summing a set of elementary grating waveforms. 

Borrowing their terminology, if we wish to generate $N$ beams, indexed by $m$, at positions $(x_m,y_m)$ in the atom plane, we write an individual SLM grating wavevector as 
\begin{align}
\mathbf{k}_m=\frac{2\pi M}{\lambda F}(x_m,y_m)\ ,
\end{align}
where $F$ is the focal length of the final lens before the cavity array and $M$ is the magnification of the telescope mapping the SLM plane to that lens. We then write the SLM phase mask as 
\begin{align}
\Phi_{\mathrm{SLM}}(\mathbf{r}) = \arg \left( \sum_m a_m e^{i(\mathbf{k}_m \cdot \mathbf{r} + \theta_m)} \right)\ ,
\end{align}
where $\mathbf{r}$ is the SLM plane coordinates, and $a_m$ and $\theta_m$ are free amplitudes and phases used during array homogenization. The intensity in the atom plane is then given as
\begin{align}
I_{\mathrm{OT}} = \left| \mathcal{F}(A_{\mathrm{SLM}}) \right|^2\ ,
\end{align}
where \mbox{$A_{\mathrm{SLM}}=\sqrt{I_0}\exp(i \Phi_{\mathrm{SLM}})$}, $\mathcal{F}$ is the Fourier transform, and $I_0$ is the beam intensity incident on the SLM. By tuning the $(x_m,y_m)$ parameters, we can achieve precision independent control of individual beam positions. 

We typically initialize a regular array with approximately the correct array spacing, and then run a positional optimization routine where we turn on one output beam of the SLM at a time, and optimize its corresponding $(x_m,y_m)$ values in order to maximize the height of the cavity reflection signal as a proxy for the incoupling. We generally find that incoupling can be improved by up to $20\%$ using this technique. Incoupling can be further improved by optimizing corrective global waveforms based on sums of Zernike polynomials, though we do not typically use this technique. Once the array is initialized we then use WGS optimization of the $a_m$ and $\theta_m$ parameters in order to homogenize the array~\cite{chew2024ultraprecise}. After a few rounds of feedback, with the depth measured via an atomic signal as an optimization target, we achieve a homogeneity (standard deviation over mean) of the trap depth of ${\approx}7\%$ (Ext. Data Fig.~\ref{efig:depth}). Note that this value is dependent on both the intensity homogeneity and the incoupling homogeneity, which are not entirely decoupled because of crosstalk and the WGS optimization. For now we leave a careful study of the optimal algorithm for reaching even better homogeneity to future work.

\subsection{Next-generation cavity design}
\label{met:nextgen}
Here we summarize some pertinent information about the next-generation cavity design showcased in Fig.~\ref{fig:nextgen}, with further details in Ref.~\cite{soper2025cavity}.

The cavity is essentially an ``unfolded'' version of the first-generation cavity design, where we replace the in-vacuum curved mirror with a second 4f telescope and planar end mirror. Both planar mirrors have $99.9\%$ reflectivity. Spherical lenses are made by Thorlabs and V-coated by Layertec. Aspheric lenses are a custom shape optimized to minimize higher-order coupling and resultant clipping loss; they are manufactured by Asphericon and broadband coated by EssentOptics to account for the high AOI of input beams. Both aspheres have focal lengths of 10 mm, with a $\approx5.7$ mm working distance. The MLA is stock from Edmund Optics (21-153) with a $30\times30$ array of lenslets with $300\ \upmu$m pitch and a 4.8 mm focal length. Internal losses per roundtrip are estimated from the finesse at $\approx5.5\%$. For further cavity parameters and expected performance, see Ext. Data Table~\ref{etab:parameters}.

In practical applications in order to collect photons from the cavity, one of the planar end mirrors would have an increased transmissivity~\cite{shadmany2025cavity}. For a maximum finesse of 110, we numerically find the optimal outcoupler reflectivity would be $82\%$, which would result in a $55\%$ cavity collection efficiency with a effective finesse of 25 - the corresponding linewidth is the relevant quantity for evaluating how many cavities are simultaneously degenerate, which in our case is greater than 400. In a far stricter, and less practically relevant analysis, we find that greater than 140 cavities are degenerate to within an effective finesse of 110. We believe the degeneracy can be greatly improved by at least several factors by improving alignment of the intra-cavity lenses to eliminate astigmatic effects, which we leave to future work.

\subsection{Data statistics and analysis}
\label{met:datastatistics}
In Fig.~\ref{fig:schematic}, the fluorescence image is an average over ${\approx}1500$ images. In Fig.~\ref{fig:imaging}, histograms are aggregated over ${\approx}2000$ shots. When calculating the discrimination fidelity we apply thresholding and Gaussian smoothing techniques~\cite{bergschneider2018spinresolved} to reduce camera noise (Ext. Data Fig.~\ref{efig:processed}); such methods are common in atom array experiments working with low signal counts~\cite{scholl2023erasure,ma2023highfidelity,su2025fast}. Finesse and degeneracy metrics reported in Figs.~\ref{fig:cooperativity} and~\ref{fig:nextgen} are averaged over a few dozen measurements for each cavity.

Correlations in Figs.~\ref{fig:schematic},~\ref{fig:imaging}, and~\ref{fig:fiberarray} are all Pearson correlation coefficients, defined as 
\begin{align}
\rho_{X,Y}=\frac{\textrm{cov}(X,Y)}{\sigma_x\sigma_y}\ ,
\end{align}
where $X$ and $Y$ are respective datasets of imaging counts, $\textrm{cov}$ is their covariance, and $\sigma_x$ and $\sigma_y$ are their individual standard deviations.

In Fig.~\ref{fig:fiberarray} we read out an array of four cavities using a four-channel fiber array (Precision Micro-Optics) coupled to a set of SPCMs (Excelitas Technologies). We did not have access to four identical SPCMs for this work, and so utilized $2\times$ SPCM-AQRH-14-FC, $1\times$ SPCM-AQRH-14 with a homemade fiber coupler, and $1\times$ SPCM-AQRH-13-FC. The SPCM-AQRH-14 and SPCM-AQRH-13-FC had noticeably worse background counts than the other modules, and so when analyzing the SPCM traces in the main text we shift and rescale all counts to be in the same relative range. In particular, for each histogram in Fig.~\ref{fig:fiberarray} of the main text we first fit with a double-Gaussian model, then shift each histogram such that the mean of the `no-atom' peak is at 0, and then rescale all counts by the optimal discrimination threshold, such that any normalized counts above 1 correspond to an atom being present, and any normalized counts below 1 correspond to no atom. Histograms in Fig.~\ref{fig:fiberarray}d are aggregated over ${\approx}2000$ shots each with 150 ms of imaging time discretized into 15 bins which are 10 ms long. For the time-series in Fig.~\ref{fig:fiberarray}e we apply a 10 ms equal weight moving sum, with 1 ms underlying bins.

\subsection{Scaling while maintaining degeneracy}
\label{met:degeneracy}
In Ext. Data Fig.~\ref{efig:sensitivity} (and Fig.~\ref{fig:novelty} of the main text) we show that multiple cavities can be made simultaneously degenerate by changing the position of the intra-cavity spherical lens into a 4f configuration. We find that cavity frequency is linearly sensitive to spherical lens displacement, with a slope that scales quadratically with cavity position relative to the telescope axis:
\begin{align}
\Delta_f=\frac{x^2}{\xi}\Delta_z\ ,
\end{align}
where $\Delta_f$ is the frequency shift of the cavity, $\Delta_z$ is the displacement of the lens, $x$ is an integer cavity index, and $\xi=36(3)$ mm/FSR is a fitted coefficient (Ext. Data Fig.~\ref{efig:sensitivity}). To find the maximum array size, that can be made simultaneously degenerate, we set $\Delta_f\approx 1/F$, where $F$ is the cavity finesse, and $\Delta_z=z_0$, the minimum accuracy of positioning the lens. We then solve for the maximum array size (assuming a circular array) of
\begin{align}
\label{eq:arraysize}
N\approx\frac{\pi\xi}{F z_0}\ ,
\end{align}
which we plot in Ext. Data Fig.~\ref{efig:sensitivity}.

We see that for $F\approx100$, the positioning accuracy must be maintained at a micron level to maintain degeneracy across thousands of sites. In Fig.~\ref{fig:nextgen} we showcase a few hundred modes being maintained at degeneracy through careful passive lens positioning with a piezo mount. To realize even more precise control, we imagine locking the length of the cavity microscope array (using the planar end-mirror, as we do in this work) to the cavity resonance of a single central cavity mode, and then locking the position of the spherical lens (mounted on a piezo stage) to the resonance of a far-offset cavity mode. 

We strongly note that further improvements could be made by studying how $\xi$ could be maximized through choice of the lens focal length, the telescope demagnification, etc. For instance, we imagine that replacing the spherical lens with a custom asphere would greatly reduce spherical aberrations, and thus might increase $\xi$; we leave such investigations for future work.

\subsection{Photon loss budget and projected readout rates}
\label{met:photonbudget}
The photon collection rate sets the speed with which high-fidelity single atom detection can be performed. We enumerate the efficiencies of our system in Ext. Data Table~\ref{etab:collection}, and provide more details below, beginning by describing losses due to intra-cavity elements.

\subsubsection{Internal cavity losses}
\label{met:internallosses}
Given the cavity finesse, $F$, we calculate the total cavity losses per round-trip, $\rho$, as
\begin{align}
F = \frac{\pi}{2 \arcsin\left( \frac{1 - \sqrt{1-\rho}}{2 \sqrt[4]{1-\rho}} \right)}\ .
\end{align}
Using the finesse measurement in Fig.~\ref{fig:cooperativity} of the main text, this yields an estimated total loss of $\rho=0.373(12)$, corresponding to an internal loss (which removes the intentional loss coming from the outcoupling mirror) of $\rho_0=0.347(13)$. A similar measurement and calculation for a different $90\%$ reflectivity outcoupler (not shown) yields an internal loss of $\rho_0=0.327(23)$. Averaging both measurements gives a best-estimate internal loss of $\rho_0=0.337(13)$.

We compare these numbers versus independent estimates of the per-element loss from various out-of-vacuum and in-vacuum measurements (Ext. Data Table~\ref{etab:losses}). Such independent estimates show a much higher loss from the curved spherical mirror than expected from the stated coating, which we surmise could be due to the high NA of the beam and the fact that the mirror introduces spherical aberrations that likely scatter into higher order cavity modes which are eventually clipped. Other significant losses come from the relative misalignment of the angle between the curved mirror and the asphere incurred during gluing and insertion into vacuum. Per-element losses multiply to a total internal loss of $0.25(2)$, with the difference versus $\rho_0$ above attributed to slough during integration and hand-alignment of all optics together.

\begin{table}[t]
    \centering
    \caption{\textbf{Cavity internal losses.} Individual optic losses are listed per pass of the light, alongside the number of passes through that optic during one cavity round-trip. Values with error bars are measured, values without error bars are manufacturer specifications. Total loss is calculated in two ways, from full-system finesse measurements, and element-wise from the individual losses of intra-cavity optics. The element-wise estimate is calculated as $1-\prod_i (1-l_i)^p$, where $l_i$ is the component loss, and $p$ is the number of passes. Note that loss could either be due to absorption, clipping, or scattering into higher-order modes of the cavity.}
    \vspace{2mm}
    \begin{tabular}{l l c}
        \toprule
        \textbf{Loss source} & \textbf{\%} & \textbf{Passes} \\
        \midrule
        Microlens array               & 0.5 & 4\\
        Spherical lens           & 0.25 & 4\\
        Vacuum chamber window & 0.5(1) & 4 \\
        Aspheric lens           & 0.6(1) & 4\\
        Curved mirror           & 1.8(4) & 2\\
        Mirror-asphere misalignment\ \     & 4.3(5) & 4 \\
        \midrule
        \textbf{Total (estimated element-wise)}    & \textbf{25(2)} & \\
        \textbf{Total (estimated from finesse)\ \ \ }   & \textbf{33.7(1.3)} &  \\

        \bottomrule
    \end{tabular}
    \label{etab:losses}
\end{table}

\subsubsection{Cavity collection}
\label{met:cavitycollection}
The percent of photons that are scattered into the cavity is given by $C/(1+C)$, where $C$ is the cooperativity. Due to internal losses, a photon which is scattered into the cavity may still be lost inside the resonator before it leaks out to the detector, meaning the percent of photons exiting the cavity through the outcoupler is \mbox{$P_\textrm{col}=\Lambda C/(1+C)$}; we typically write \mbox{$\Lambda=L_\textrm{out}/(L_\textrm{out}+L_\textrm{int})$}, where $L_\textrm{out}$ is the transmission of the relevant output port (which in our case includes both output ports on the outcoupling mirror) and $L_\textrm{int}$ includes all other internal losses (Ext. Data Table~\ref{etab:losses}), but this only holds in the limit of small losses, i.e. $L_\textrm{out},L_\textrm{int}\ll1$. In this first edition of the cavity array microscope, our losses are not in this regime, so we derive the full expression for $\Lambda$.

After a photon is emitted into the cavity mode, it traverses one quarter round-trip of the cavity mode (from the atoms to the outcoupling mirror). It then continues making round-trips in the cavity until it is lost either from (nonideal) internal losses or through the outcoupling mirror. We may write the sequence of events the photon sees as \mbox{$\mathbb{S}=I_0M_1I_1I_2M_2I_3I_4M_3I_4I_5\cdots$}, where $I$ is an interaction with the internals of the cavity during a quarter round-trip, and $M$ is an interaction with the outcoupling mirror. On each interaction with the mirror, the probability to be lost is given by the transmission of the mirror, in our case $P_M=0.1$. During each quarter round-trip the probability to be lost internally is approximately $P_I=1-(1-\rho_0)^{1/4}=0.098(4)$. We then ask the question: during the sequence $\mathbb{S}$, what is the probability the photon is lost during any $M$ event?

We first consider the $M_n$ event. Before, there are $n-1$ $M$ events, and there are $2n-1$ $I$ events, meaning the probability to be lost at the $M_n$ event is \mbox{$(1-P_I)^{2n-1}(1-P_M)^{n-1}P_M$}. The total probability to lose the photon at any $M$ event is then
\begin{align}
\Lambda&=\sum_{n=1}^\infty(1-P_I)^{2n-1}(1-P_M)^{n-1}P_M\nonumber\\
&=P_M(1-P_I)\sum_{n=0}^\infty\Big((1-P_M)(1-P_I)^2\Big)^n\nonumber\\
\label{eq:mirrorloss}
&=\frac{P_M(1-P_I)}{1-(1-P_M)(1-P_I)^2}\ .
\end{align}
Note that in the limit that $P_M,P_I\ll1$, Eq.~\eqref{eq:mirrorloss} simplifies to \mbox{$\Lambda\approx P_M/(P_M+2P_I)\approx L_\textrm{out}/(L_\textrm{out}+L_\textrm{int})$}, as expected.

As such, the overall cavity collection efficiency is given by
\begin{align}
    P_\textrm{col} = \Big(\frac{C}{1+C}\Big) \Big(\frac{P_M(1-P_I)}{1-(1-P_M)(1-P_I)^2}\Big)\ ,
\end{align}
which for our parameters yields $P_\textrm{col}=18.1(1.5)\%$. This is the peak collection, but it is further reduced to $13.9(1.1)\%$ when including corrections from averaging over the positional spread of the atom along the cavity axis due to finite temperature (assumed to be $\approx 25\ \upmu$K, with a trap frequency of $\approx280$ kHz, Ext. Data Fig.~\ref{efig:depth}), as well as a factor to account for the fact we currently do not perform optical pumping, and so photons are emitted with approximately random polarization. 

After exiting the cavity, further losses include transmission through a Glan-Taylor polarizer (needed to homogenize trap polarizations due to limited selection of dichroics, described above), telescopic optics, and the quantum efficiency of our Cascade II Photometrics 512 EMCCD, leading to an overall imaging efficiency of $4.6(4)\%$ (Ext. Data Table~\ref{etab:collection}). 

\begin{table}[t]
    \centering
    \caption{\textbf{Imaging efficiency budget.}}
    \vspace{2mm}
    \begin{tabular}{l c }
        \toprule
        \textbf{Source} & \textbf{Efficiency (\%)} \\
        \midrule
        Cavity               & 13.9(1.1) \\
        Glan-Taylor           & 46 \\
        \quad \quad From polarization & 50 \\
        \quad \quad From transmission & 92 \\
        Telescopes to camera           & 96 \\
        Quantum efficiency           & 75 \\
        \midrule
        \textbf{Total}          & 4.6(4) \\

        \bottomrule
    \end{tabular}
    \label{etab:collection}
\end{table}

\subsubsection{Improving photon collection rates}
\label{met:improvingphoton}
There are multiple pathways to improve photon collection rates in the current apparatus. The most impactful near-term gain will come from using a proper dichroic to obviate the need for the polarizer, yielding a $2.2\times$ increase in collection. Further gains of a few tens of percent each could be achieved by: 1) optically pumping atoms to drive with a well-defined polarization, 2) better localizing the atom via improved cooling and transferring from the pseudo-dipole trap into a lattice potential with $2\times$ increase in longitudinal trapping frequency~\cite{shadmany2025cavity}, 3) using a more optimal $75\%$ reflectivity outcoupling mirror to better match our current internal loss rates. These improvements would lead to an atom-to-camera imaging efficiency of $17(1)\%$, almost a $4\times$ improvement without adjusting any intra-cavity optics. 

From there, further improvements are readily available by optimizing our optical molasses driving the atomic fluorescence: when the collection efficiency is higher, imaging can be made shorter and so the scattering rate can be increased without affecting the imaging survival probability. For instance, in our first paper working with a single atom in a single lens-based cavity~\cite{shadmany2025cavity}, we used a roughly $5\times$ higher scattering rate, enabling imaging in $130\ \upmu$s with high survival and high fidelity; the lower scattering rate employed here was chosen to ensure high survival in the long-imaging time/high-survival limit, and was not specifically re-optimized for the short-time regime. Using this increased scattering rate, along with our improved collection efficiency above, we estimate we could perform imaging in ${\approx}200\ \upmu$s across the entire array. By using more advanced stroboscopic imaging techniques, such as in Ref.~\cite{bluvstein2024logical}, we estimate at least a further ${\approx}20\%$ gain is achievable. 

Finally, substantial further gains are likely in our next-generation version of the cavity array microscope with a mean finesse of $F=110$, corresponding to a total internal loss of around $5\%$ (around $6\times$ lower than our current value). An atom-to-camera imaging efficiency of $45\%$ is then feasible, implying as low as ${\approx}60\ \upmu$s imaging across the entire array, with room to decrease further as the finesse continues to improve and as more advanced cavity detection methods are discovered. Realizing these gains will not only improve the practical utility of the platform, but will also validate the strong light-matter coupling we achieve, alongside more direct tests such as cavity-resolved vacuum Rabi splitting measurements which we will undertake following suitable technical improvements, e.g. implementation of single-site addressing systems.

\subsection{Cavity mode waist and stability region}
\label{met:modewaist}
We computationally determine the mode waist of our cavity via conventional ABCD matrix calculations, using the eigenvalues and vectors of the round-trip, complex $q$ parameter. Typical paraxial ABCD calculations require all optical elements to be aligned along a central axis. However, in our cavity array, the central axes of the spherical lens, aspheric lens, and curved mirror do not line up with the central axes of any microlens in the microlens array. To make our cavity compatible with paraxial ABCD matrix calculations, we align all optics on a common central axis. This centrally-aligned cavity differs from the cavities in the array because its mode returns to itself after one round-trip, whereas the off-axis cavities in the cavity array have an inversion around the curved mirror which causes them to take two round-trips before coming back on themselves. Therefore, we use an ABCD matrix that includes two round-trips through the centrally-aligned cavity. 

Using this ABCD matrix, we compute the cavity stability diagram as a function of the displacement of each of the cavity optics (Ext. Data Fig.~\ref{efig:abcdcalc}). For our most sensitive optic, the in-vacuum asphere, we sit near the top of the stability diagram where we calculate a waist of 1.08~$\upmu$m, in good agreement with our experimentally measured average waist across the array. 

At the center of the stability diagram, the expression for the mode waist in the atom plane is approximately
\begin{align}
w_{0}=\frac{1}{M}\times\sqrt{\frac{f_{MLA}\lambda}{\pi}},
\end{align}
as long as $f_{MLA}^2\ll M^4ROC^2$. This expression arises from the MLA-end mirror distance which forms a half-planar confocal cavity. The mode size at the microlens is equal to the mode size at the mirror of the equivalent confocal cavity, which then gets de-magnified by a factor of $M$ from the microlens plane into the atom plane. The fact that the mode waist is the same for all cavities arises because the telescope has the same action on all rays. Each microlens defines an independent half-planar confocal cavity, and since the telescope acts equally on all rays after the microlens, all cavities have identical mode waist and Gouy phase, up to aberrations, misalignment, or imperfections of the optics. The width of the stability region (how far the asphere can be displaced along the cavity axis while maintaining a finite mode waist) is $\Delta z=2\times \frac{\pi w_0^2}{\lambda}$, which is twice the Rayleigh range of the waist at the stability region center, analogous to the case of a standard two-mirror resonator.

There is one subtlety to the statement that all rays after the microlens are acted on equally. This is true of the 4f telescope, but not of the curved mirror, which forms a 1:1 2f imaging system. For an input ray, $(d,s)$ the 2f system will map it to $(-d,s+\Delta s)$ where $\Delta s=\frac{2d}{R}$. This small error in the slope will not change the mode size, however it could destabilize the off axis cavities. To determine if this perturbation is tolerable, we compare it to the divergence angle of the cavity mode at the microlens, which is the largest angle that can be stabilized. The angular error from the curved mirror propagated to the microlens array is $\Delta\theta=\frac{1}{M}\arctan{\frac{2d}{R}}$. For a $5$~$\upmu$m displacement in the atom plane, we find that it is 330$\times$ smaller than the confocal divergence angle, and thus we expect most of the light to remain trapped stably inside of the resonator. While this could become a limiting factor at high finesses or further out cavities, we do not find it to be limiting us in this work.

\subsubsection{Ray tracing simulations}
\label{met:raytracing}
For the simulations in Fig.~\ref{fig:novelty}a of the main text, we use an in-house open source numerical ray tracing package~\cite{Palm_sloppy} designed to perform semi-classical analysis of novel cavity geometries~\cite{jaffe2021aberrated}. The code performs exact, non-paraxial propagation of rays through spherical, freeform aspheric optics as well as the microlens array. Coupling of paraxial ABCD calculations and full ray-tracing allow for efficient finding of eigenrays as well as extraction of effective off-axis ABCD matrices. To study the local stability and finesse of the off-axis cavities, input beams are generated in the transverse plane with zero slope and propagated through the full system for up to 100 round-trips. Since we are interested in the additional loss from aberrations and clipping, we assume all mirrors are perfectly reflective, and all lenses are perfectly transmissive. The number of round-trips before a ray clips is then a proxy for the finesse of a mode corresponding to the guide ray. When studying the MLA cavity geometry, we use the exact parameters and spacings used in the experimental system, but when studying the no-MLA geometry, we adjust spacings to make the equivalent cavity stable and with a comparable waist size for the central mode for conservative comparison.

\clearpage

\end{document}